\documentclass[11pt,a4paper]{article}

\usepackage{graphicx}
\usepackage{xcolor}
\usepackage[caption=false]{subfig}
\usepackage{mathrsfs,mathtools}
\usepackage{physics,amssymb}
\usepackage{siunitx}
\usepackage{bm}
\usepackage{braket}
\usepackage{listings}
\DeclareMathOperator{\sinc}{sinc}
\usepackage{cases}
\usepackage{comment}
\usepackage{soul}
\usepackage{cancel}
\usepackage{cases}
\usepackage[utf8]{inputenc}
\usepackage{url}
\usepackage{float}
\usepackage{longtable}
\usepackage[normalem]{ulem}
\usepackage{xspace}
\usepackage{aas_macros}
\setstcolor{red}
\usepackage{jcappub}
\hypersetup{colorlinks=true
,urlcolor=DARKBLUE
,anchorcolor=DARKBLUE
,citecolor=DARKBLUE
,filecolor=DARKBLUE
,linkcolor=DARKBLUE
,menucolor=DARKBLUE
,linktocpage=true
,pdfproducer=medialab
}

\newcommand{\ee}{\mathrm{e}}

\newcommand{\fNL}{f_\mathrm{NL}}

\newcommand{\MS}{\mathrm{MS}}

\newcommand{\PBH}{\mathrm{PBH}}
\newcommand{\DM}{\mathrm{DM}}

\newcommand{\tot}{\mathrm{tot}}

\newcommand{\uA}{\mathrm{A}}
\newcommand{\calC}{\mathcal{C}}
\newcommand{\uc}{\mathrm{c}}

\newcommand{\uF}{\mathrm{F}}

\newcommand{\uG}{\mathrm{G}}
\newcommand{\ug}{\mathrm{g}}

\newcommand{\bfk}{\mathbf{k}}

\newcommand{\um}{\mathrm{m}}
\newcommand{\uN}{\mathrm{N}}

\newcommand{\calO}{\mathcal{O}}

\newcommand{\calP}{\mathcal{P}}

\newcommand{\bfx}{\mathbf{x}}

\newcommand{\bae}[1]{\begin{align} #1 \end{align}}

\newcommand{\bfe}[4]{
\begin{figure} 
	\centering
	\includegraphics[#1]{#2}
	\caption{#3}
	\label{#4}
\end{figure}}
\newcommand{\beae}[1]{\begin{equation}\begin{aligned} #1 \end{aligned}\end{equation}}

\newcommand{\bme}[1]{\begin{multline} #1 \end{multline}}

\definecolor{MONZA}{HTML}{CF000F}
\definecolor{DARKBLUE}{HTML}{00008b}
\definecolor{DARKMAGENTA}{HTML}{8b008b}
\definecolor{DARKCYAN}{HTML}{008B8B}
\definecolor{DARKORANGE}{HTML}{FF8C00}

\title{Simulation of Primordial Black Holes with large negative non-Gaussianity}

\author[a]{Albert Escriv\`a,}
\author[b,c]{Yuichiro Tada,}
\author[d,e]{Shuichiro Yokoyama,}
\author[c]{and Chul-Moon Yoo}

\affiliation[a]{Service de Physique Th\'eorique, Universit\'e Libre de Bruxelles, \\ 
Boulevard du Triomphe CP225, B-1050 Brussels, Belgium}
\affiliation[b]{Institute for Advanced Research, Nagoya University, \\
Furocho Chikusaku Nagoya, Aichi 464-8601 Japan}
\affiliation[c]{Department of Physics, Nagoya University, \\
Furocho Chikusaku Nagoya, Aichi 464-8602 Japan}
\affiliation[d]{Kobayashi Maskawa Institute, Nagoya University, \\ 
Chikusa, Aichi 464-8602, Japan}
\affiliation[e]{Kavli IPMU (WPI), UTIAS, The University of Tokyo, \\ 
Kashiwa, Chiba 277-8583, Japan}

\emailAdd{albert.escriva@ulb.be}
\emailAdd{tada.yuichiro.y8@f.mail.nagoya-u.ac.jp}
\emailAdd{shu@kmi.nagoya-u.ac.jp}
\emailAdd{yoo@gravity.phys.nagoya-u.ac.jp}

\abstract{In this work, we have performed numerical simulations of 
primordial black hole (PBH) formation in the Friedman--Lemaître--Robertson--Walker universe filled by radiation fluid, introducing the local-type non-Gaussianity to the primordial curvature fluctuation.  
We have compared the numerical results from simulations with previous analytical estimations on the threshold value for PBH formation done in the previous paper~\cite{Kitajima:2021fpq}, particularly for negative values of the non-linearity parameter $\fNL$. 
Our numerical results show the existence of PBH formation of (the so-called) type I also in the case $\fNL \lesssim -0.336$, which was not found in the previous analytical expectations using the critical averaged compaction function. 
In particular, although the universal value for the averaged critical compaction function $\bar{\calC}_{c}=2/5$ found previously in the literature is not satisfied for all the profiles considered in this work, an alternative direct analytical estimate has been found to be roughly accurate to estimate the thresholds, which gives the value of the critical averaged density with a few~\% deviation from the numerical one for $f_{\rm NL}\gtrsim -1$.}

\begin{document}

\maketitle
\flushbottom

\section{Introduction}

Primordial black holes (PBHs) may have been formed in the very early universe due to high and rare peaks on the distribution of density perturbations~\cite{hawking1,hawking2,acreation1}. 
If those high peaks were sufficiently large, they could have undergone gravitational collapse and formed black holes. 
The research field of PBHs has become very active nowadays thanks to the development of gravitational wave (GW) astronomy, in particular, due to the first direct gravitational wave detection of binary black hole merger~\cite{LIGO}, as massive PBH binaries may be the source of such GW events possibly~\cite{Bird,Clesse:2016vqa,Sasaki:2016jop}. 
PBHs can be also the candidate of dark matter in our universe if their mass is as light as asteroids (see, e.g., Refs.~\cite{Carr1,darkmatter1,darkmatter2,darkmatter3,darkmatter4,darkmatter5,darkmatter6,Bird,darkmatter8,Clesse:2016vqa,Clesse:2015wea,Clesse:2018ogk,Tada:2019amh,Atal:2020yic,Atal:2020igj,Bartolo:2018rku,Clesse:2017bsw,Ezquiaga:2019ftu}).

Note that there have been suggested many other mechanisms for PBH formation~\cite{Carr1},
but we focus on PBHs formed by the collapse of curvature fluctuations in the radiation-dominated universe in this paper (c.f., see Refs.~\cite{Passaglia:2021jla,Yoo:2021fxs} for PBHs from isocurvature and Ref.~\cite{Harada:2016mhb} for PBH formation in the matter-dominated universe).
In this case, the abundance of those black holes depends exponentially on the threshold of their formation~\cite{carr75}. The threshold value for PBH formation is the minimum amplitude of the cosmological perturbation in such a way that the perturbation collapses and forms a black hole. The threshold is not a universal quantity
for a fixed equation of state but it depends on the specific details of the shape of the curvature fluctuation~\cite{musco2005,hawke2002,Harada:2015yda,Niemeyer2,Shibata:1999zs,Nakama:2014fra,Musco:2018rwt,Escriva:2019nsa}.

In principle, numerical simulations are needed for an accurate determination of the threshold~\cite{musco2005,hawke2002,Harada:2015yda,Niemeyer2,Shibata:1999zs,Nakama_2014,Musco:2018rwt,Escriva:2019nsa,Escriva:2021aeh}. 
Nevertheless, some useful analytical estimations have been pointed out in the literature~\cite{carr75,harada,Escriva:2020tak,Escriva:2019phb}. 
In particular, the ones in Refs.~\cite{Escriva:2020tak,Escriva:2019phb} take into account the shape of the curvature profile. These analytical profile-dependent estimations were based on the use of the averaged critical compaction function, which is a quantity that seems approximately universal over several profiles, only depending on the equation of state of the fluid filling the universe~\cite{Escriva:2019phb}. Specifically, for the case of PBH formation in the Friedman--Lemaître--Robertson--Walker (FLRW) universe
filled with radiation, the critical value is given by $2/5$~\cite{Escriva:2019phb}. 
In addition, there a certain fitting formula for a general profile has been proposed with a single dimensionless fitting parameter $q$. Combining this analytic fitting with the average compaction function approach, an analytic estimation for the PBH threshold has been established, which would be quite useful for the statistical prediction of the PBH abundance as the threshold can be parametrised for a wide class of the perturbation profile.
An example is shown in Ref.~\cite{nonlinear}.

On the other hand than the threshold issue itself, 
the precise statistical estimation method of the PBH abundance has been developed in several ways~\cite{nonlinear,DeLuca:2019qsy,newnicolla,Erfani:2021rmw,Wu:2020ilx,DeLuca:2020ioi,Young:2020xmk,Yoo:2020dkz,yoo,Gow:2020bzo,Young:2020xmk,Young:2019osy,Young:2019yug,Young:2013oia,Young:2014ana,Germani:2018jgr,Garriga,Suyama:2019npc,Ando:2018qdb,Zaballa:2006kh,Yokoyama:1998xd,Tada:2021zzj}. 
Particularly in the so-called peak theory, the typical profile of high peaks of the curvature fluctuation (and thus whether a PBH is formed or not) can be statistically discussed with the power spectrum if the curvature field follows the Gaussian statistics. 
However if the curvature fluctuation shows non-Gaussianity, the PBH abundance is expected to change substantially~\cite{Atal:2018neu,vicente-garriga,Hayato2,Cai:2017bxr,Bullock:1996at,Pattison_2017,PinaAvelino:2005rm,Young:2013oia,Young:2014ana,Young:2014oea,Riccardi:2021rlf,Young:2015cyn,yoo,Hidalgo:2007vk,Atal:2021jyo,Kitajima:2021fpq,Davies:2021loj,Taoso:2021uvl}.\footnote{Some papers (e.g., Ref.~\cite{Young:2022phe}) claim that the non-Gaussian effects may be weaker, though.} 
Recently, Ref.~\cite{Kitajima:2021fpq} studied the case of the local-type non-Gaussianity (parametrised, e.g., by the non-linearity parameter $\fNL$), making the direct use of the critical averaged compaction function (i.e., without the fitting formula by the $q$ parameter). As an interesting remark, according to the criterion that the critical value of the averaged compaction function is 2/5, there no type I PBH 
\footnote{Following Ref.~\cite{Kopp:2010sh}, we divide super-horizon fluctuations into type II and I according to whether they have the region in which the areal radius is a decreasing function of a radial coordinate or not, respectively (see Eq.~\eqref{eq:conditiontype2} and surrounding descriptions). PBH formations can be also classified into type II and I depending on the type of the initial fluctuation. }
has been found for some range of negative values in $\fNL$ (precisely, $\fNL\lesssim-0.336$), somehow against the intuition.

The main aim of this work is to test the formation of PBHs (type I, specifically) particularly in the case of negatively large $\fNL$, with use of full numerical simulations. 
Such a numerical study has been already done in Ref.~\cite{Atal:2019erb} for positive $\fNL$, but not for negative $\fNL$ due to the difficulty of such simulations. In our work, we have used a substantially improved numerical code (in comparison with the one used in Ref.~\cite{Atal:2019erb}) to be able to handle such profiles. We indeed found type I PBH formation even for $\fNL\lesssim-0.336$, in contrast to the average compaction function approach. Despite this failure of the average compaction, the fitting approach with the $q$ parameter somehow works well up to $\fNL\sim-1$, beyond which the $\fNL$ expansion itself may be doubtful.
We further show examples of the predicted PBH mass spectra, the current PBH abundance in terms of their mass.

The rest of the paper is organised as follows. We first review the basics of the peak theory, including the local-type non-Gaussianity in Sec.~\ref{sec: Peak profile with the local-type non-Gaussianity}, and then the fitting formula with the $q$ parameter in Sec.~\ref{sec:analitical}. The initial conditions and set up are described in Sec.~\ref{sec: Initial conditions and set up for PBH formation}, and the main results of simulations are shown in Sec.~\ref{sec: Numerical results and comparison with analytical estimations} with a comparison to the $q$ parameter approach. The PBH mass function is discussed in Sec.~\ref{sec:mass}. Sec.~\ref{sec: Summary and conclusions} is devoted to summary and conclusions. 
We use in this work geometrised units with $c=G=1$.

\section{Peak profile with the local-type non-Gaussianity}
\label{sec: Peak profile with the local-type non-Gaussianity}

In this section, we summarise the peak profile of the primordial curvature perturbation including the local-type non-Gaussian correction.
Let us first review the statistics of the Gaussian field $\zeta_\ug$, following, e.g., Ref.~\cite{Bardeen:1985tr}.
The Gaussian field is characterised only by its power spectrum,
\bae{
    \braket{\zeta_\ug(\bfk)\zeta_\ug(\bfk^\prime)}=\frac{2\pi^2}{k^3}\calP_\ug(k)(2\pi)^3\delta^{(3)}(\bfk+\bfk^\prime),
}
where $\zeta_\ug(\bfk)$ is the Fourier mode of $\zeta_\ug$.
Particularly, it is known that a local high peak of such a Gaussian field typically takes a spherically symmetric configuration. Its profile is hence characterised by the spherically-symmetric real-space two-point function,
\bae{
    \psi(r)=\frac{1}{\sigma_0^2}\int\calP_\ug(k)\sinc(kr)\frac{\dd{k}}{k},
}
where $\sinc(z)=\sin(z)/z$ is the sinc function and $\sigma_0^2$ is the variance of $\zeta_\ug$ as
\bae{
    \sigma_0^2=\int\frac{\dd{k}}{k}\calP_\ug(k).
}
Throughout this paper, we focus on the monochromatic spectrum given by
\bae{\label{eq: monochromatic P}
    \calP_\ug(k)=\sigma_0^2k_*\delta(k-k_*).
}
In this case, the typical profile of $\zeta_\ug$ is much simplified as
\bae{\label{eq: mu psi}
    \zeta_\ug(r)=\mu\psi(r)=\mu\sinc(k_*r),
}
where $\mu$ is a random parameter following the Gaussian probability density,
\bae{
    P(\mu)=\frac{1}{\sqrt{2\pi\sigma_0^2}}\ee^{-\frac{\mu^2}{2\sigma_0^2}}.
}

In general, the primordial curvature perturbation may not be simply given by a Gaussian field.
Let us then investigate a small non-Gaussian correction, supposing the local-type template parametrised by the non-linearity parameter $f_{\rm NL}$:
\bae{
    \zeta(\bfx)=\zeta_\ug(\bfx)+\frac{3}{5}\fNL\zeta_\ug^2(\bfx).
}
In a moderate non-Gaussian case $\abs{\fNL}\sim\calO(1)$, the Gaussian field $\zeta_\ug$ should take $\calO(1)$ values as well as the full field $\zeta$ in order to realise a PBH. Therefore, $\zeta_\ug$ is also understood as a ``high peak" and its profile can be assumed to well given by the typical one~\eqref{eq: mu psi}. 
The full field hence reads
\bae{
    \zeta(r)=\mu\sinc(k_*r)+\frac{3}{5}\fNL\mu^2\sinc^2(k_*r).
    \label{zeta_t}
}

Below we will study the PBH formation, giving this curvature perturbation on a superhorizon scale as an initial condition.
There, the spacetime metric can be written as a perturbed FLRW one as~\cite{Shibata:1999zs},
\bae{\label{eq:metric_superhorizon}
    \dd{s^2}=-\dd{t^2}+a^2(t)\ee^{2\zeta(r)}(\dd{r^2}+r^2\dd{\Omega^2}),
}
where $a(t)$ is the scale factor, which evolves as $a(t)=a_{0}(t/t_{0})^{1/2}$ in the radiation-dominated universe, and $\dd{\Omega^2} = \dd{\theta^{2}}+\sin^{2}(\theta)\dd{\phi^{2}}$ is the angular line element. 
Associated with this perturbed metric, the so-called \emph{compaction function}~\cite{Shibata:1999zs,Harada:2015yda} defined as the mass excess inside a given areal radius $R(r)=a \ee^{\zeta(r)}r$ is useful and has been investigated extensively in the literature as a criterion of the PBH formation.
In spherical symmetry, it is defined by
\bae{\label{eq:compact}
    \calC(r) = 2 \frac{M_\MS-M_\uF}{R(r)},
}
where $M_\MS=4\pi\int_0^R\rho\tilde{R}^2\dd{\tilde{R}}$ is the Misner--Sharp mass (which takes into account the kinetic and potential energy) and $M_\uF$ is the mass expected in the FLRW background, defined as $M_\uF=4 \pi \rho_{\uF}R^{3}/3$. 
$\rho$ is the energy density of the full fluid, while $\rho_\uF$ is that of the FLRW background, which evolves as $\rho_\uF = \rho_{\uF,0}(t/t_{0})^{-2}$ 
in the radiation-dominated universe.
From this definition, one notes that the compaction function can be also understood as the average of the density contrast over a given volume at the moment of horizon reentry $R=1/H$ ($H$ is the Hubble factor), i.e,
\bae{\label{eq:CC}
   \calC=(RH)^{2}\left.\pqty{4\pi\int_0^{R}\frac{\delta \rho}{\rho_{\uF}}\tilde{R}^2(r)\dd{\tilde{R}(r)}}\middle/\pqty{\frac{4\pi}{3}R^3(r)} \right. = (RH)^{2}\bar{\delta},
}
where $\bar{\delta}$ is the averaged density contrast.
The crucial point was shown in Ref.~\cite{Harada:2015yda}:
recalling the relation between the comoving density contrast and the curvature perturbation,
\bae{
    \frac{\delta\rho}{\rho_\uF}=-\frac{4(1+w)}{5+3w}\frac{1}{a^2H^2}\ee^{-5\zeta/2}\Delta\ee^{\zeta/2},
    \label{eq:dens_contrast}
}
the compaction function~\eqref{eq:compact} can be written in terms of $\zeta$ as

\bae{\label{eq:compact_superhorizon}
    {\cal C}(r) = \frac{3(1+w)}{5+3w} \left[1-(1+r\zeta'(r))^{2}\right]+O(\epsilon^3),
}
where $\epsilon$ is the parameter for the gradient expansion given by the Hubble scale divided by the perturbation length scale~(see Eq.~\eqref{eq:epsilon}). $w=p/\rho$ is the equation-of-state parameter with the pressure $p$, which is $w=1/3$ in the radiation-dominated universe. $\Delta$ is the Laplacian and the prime stands for radial derivative $\dd / \dd r$.
This expression is firstly derived in Ref.~\cite{Harada:2015yda}, and is time-independent to $O(\epsilon^2)$. 

The maximum of the compaction function is often seen as a useful criterion for the PBH formation. There, the threshold is defined by $\delta_\uc=\calC_\uc(r_\um)$ where $r_\um$ maximises the compaction function $\calC_\uc$ for a critical overdensity. 
The maximum radius $r_\um$ is found by the extremal condition $\calC_\uc^\prime(r_\um)=0$, i.e.,
\bae{\label{eq:rm}
    \zeta^\prime(r_\um)+r_\um\zeta^{\prime\prime}(r_\um)=0,
}
and this radius is understood as the length scale of the overdensity~\cite{Shibata:1999zs,Harada:2015yda}. 
The threshold $\delta_{\uc}$ was found numerically to be in the range $\delta_{\uc} \in [2/5,2/3]$ in the case $w=1/3$~\cite{Musco:2018rwt,Escriva:2019phb}.

Instead of the compaction function itself, Ref.~\cite{Escriva:2019phb} suggests the averaged compaction function, 
\bae{\label{eq: barCm}
    \bar{\calC}=\left.\pqty{4\pi\int_0^{R(r_\um)}\calC(r)\tilde{R}^2(r)\dd{\tilde{R}(r)}}\middle/\pqty{\frac{4\pi}{3}R^3(r_\um)} \right.,
}
as a convenient quantity which gives a more accurate criterion with the threshold $\bar{\calC}_\uc=2/5$ in the case $w=1/3$ (see Ref.~\cite{Escriva:2020tak} for a generalisation within the perfect fluid).
While this averaged compaction approach works well for positive non-Gaussianity ($\fNL>0$)~\cite{Atal:2019erb}, it fails to find the PBH formation condition (for a type I perturbation, strictly speaking, which is defined below) for negatively large non-Gaussianity, $\fNL\lesssim-0.336$~\cite{Kitajima:2021fpq}.
In this paper, we directly investigate the PBH formation condition in such a negatively non-Gaussian case, making use of a numerical simulation.

Before closing this section, we mention the two types of perturbations. For a smaller $\mu$, the areal radius $R(r)=a\ee^{\zeta(r)}r$ is monotonically increasing in $r$ along with the corresponding perturbation, which is called \emph{type I} and considered in the literature as standard.
On the other hand, the \emph{type II} perturbation~\cite{Kopp:2010sh} with a larger $\mu$ have the particularity that $R(r)$ is not a monotonic function, which means that
\bae{ \label{eq:conditiontype2}
    \exists r>0 \quad \text{s.t.} \quad \dv{R}{r}=a\ee^\zeta(1+r\zeta^\prime)<0.
}
Ref.~\cite{Kopp:2010sh} implies that the type II perturbation always leads to a PBH irrespectively of the value of the compaction function.
In our work, we only consider numerical simulations of type I perturbations.

\section{\boldmath The dimensionless $q$ parameter}\label{sec:analitical}

In Ref.~\cite{Escriva:2019phb}, an intriguing ``$q$" parameter is introduced to fit general peak profiles inside the maximal radius $r_\um$.
Instead of the comoving coordinate~\eqref{eq:metric_superhorizon}, Ref.~\cite{Escriva:2019phb} employs the (comoving) areal radius
\bae{
    \tilde{r}=r\ee^{\zeta(r)},
}
with which the metric is summarised as
\bae{
    \dd{s^2}=-\dd{t^2}+a^2(t)\bqty{\frac{\dd{\tilde{r}^2}}{1-K(\tilde{r})\tilde{r}^2}+\tilde{r}^2\dd\Omega^2}.
}
The curvatures $\zeta$ and $K$ are related as
\bae{
    \zeta(r)=\int^{\tilde{r}}_{\infty}\pqty{1-\frac{1}{\sqrt{1-K(\tilde{r})\tilde{r}^2}}}\frac{\dd{\tilde{r}}}{\tilde{r}},
}
and accordingly the compaction function can be simply expressed as
\bae{\label{eq: C in K}
    \tilde \calC(\tilde{r})=\calC(r(\tilde r))=\frac{3(1+w)}{5+3w}K(\tilde{r})\tilde{r}^2, 
}
where the radial coordinate $r$ is expressed as a function of $\tilde r$ as $r(\tilde r)$.

Ref.~\cite{Escriva:2019phb} then introduces the fiducial profile
\bae{\label{eq: Kq}
    K_q(\tilde{r})=\frac{5+3w}{3(1+w)}\frac{\tilde \calC(\tilde{r}_\um)}{\tilde{r}_\um^2}\ee^{\frac{1}{q}\bqty{1-\pqty{\frac{\tilde{r}}{\tilde{r}_\um}}^{2q}}},
}
with one parameter $q$. One finds that the parameter $q$ satisfies 
\bae{\label{eq: q def}
    q=-\frac{1}{4}\tilde{r}_\um^2\frac{\tilde \calC^{\prime\prime}(\tilde{r}_\um)}{\tilde \calC(\tilde{r}_\um)},
}
if the curvature $K$ (and then the compaction $\tilde \calC$ through Eq.~\eqref{eq: C in K}) is given by the fiducial profile~\eqref{eq: Kq}.
Inversely, Ref.~\cite{Escriva:2019phb} suggests that, given a general peak profile, the corresponding $q$ parameter is defined by Eq.~\eqref{eq: q def} and the profile can be well approximated by the fiducial one~\eqref{eq: Kq} with use of such a $q$ parameter.
In fact, there it is shown that several example profiles with the same $q$ have the same threshold value within $2\%$ errors compared with numerical results.

Once the peak profile is approximated by the fiducial one~\eqref{eq: Kq}, the averaged compaction~\eqref{eq: barCm} can be analytically obtained as
\bae{
    \bar{\calC}_q=\frac{3}{2}\ee^{\frac{1}{q}}q^{-1+\frac{5}{2q}}\bqty{\Gamma\pqty{\frac{5}{2q}}-\Gamma\pqty{\frac{5}{2q},\frac{1}{q}}}\tilde \calC(\tilde{r}_\um).
}
Recalling the universal criterion $\bar{\calC}_\uc=2/5$, the threshold value for the maximal compaction $\delta_\uc=\calC_\uc(r_\um)$ would be expressed by
\bae{
    \delta_\uc(q)=\frac{4}{15}\ee^{-\frac{1}{q}}\frac{q^{1-5/2q}}{\Gamma(5/2q)-\Gamma(5/2q,1/q)},
    \label{threshold_anali}
}
as a function of $q$. 
A broad profile in the compaction function ($q \rightarrow 0$) leads to the minimum threshold $\delta_{c} \rightarrow 2/5$, while a sharp profile ($q \rightarrow \infty$) leads to the maximum threshold $\delta_{c} \rightarrow 2/3$ as numerical works suggested~\cite{Musco:2018rwt,Escriva:2019phb}.
In our case, given the amplitude $\mu$ and the non-Gaussianity $\fNL$, the corresponding $r_\um$, $\tilde \calC(\tilde{r}_\um)$, and $q$ are obtained in order.
Since both sides of the equation \eqref{threshold_anali} have $\mu$-dependences as $\delta_{\uc}  \equiv \calC_{\uc}(r_\um;\fNL,\mu)$ and $q  \equiv q(\fNL, \mu)$, one can numerically obtain the $\mu_{\uc}$ by finding the value of $\mu$ such that the previous equation holds for a given $\fNL$. Note that the definition of the $q$ parameter~\eqref{eq: q def} can be rewritten in the coordinate $r$~\eqref{eq:metric_superhorizon} as
\bae{
\label{eq: q def2}
    q=-\frac{1}{4}r_\um^2\frac{\calC^{\prime\prime}(r_\um)}{\calC(r_\um)(1-\frac{3}{2}\calC(r_\um))}.
}

\begin{figure}
    \centering
    \begin{tabular}{cc}
        \begin{minipage}{0.45\hsize}
            \centering
            \includegraphics[width=\hsize]{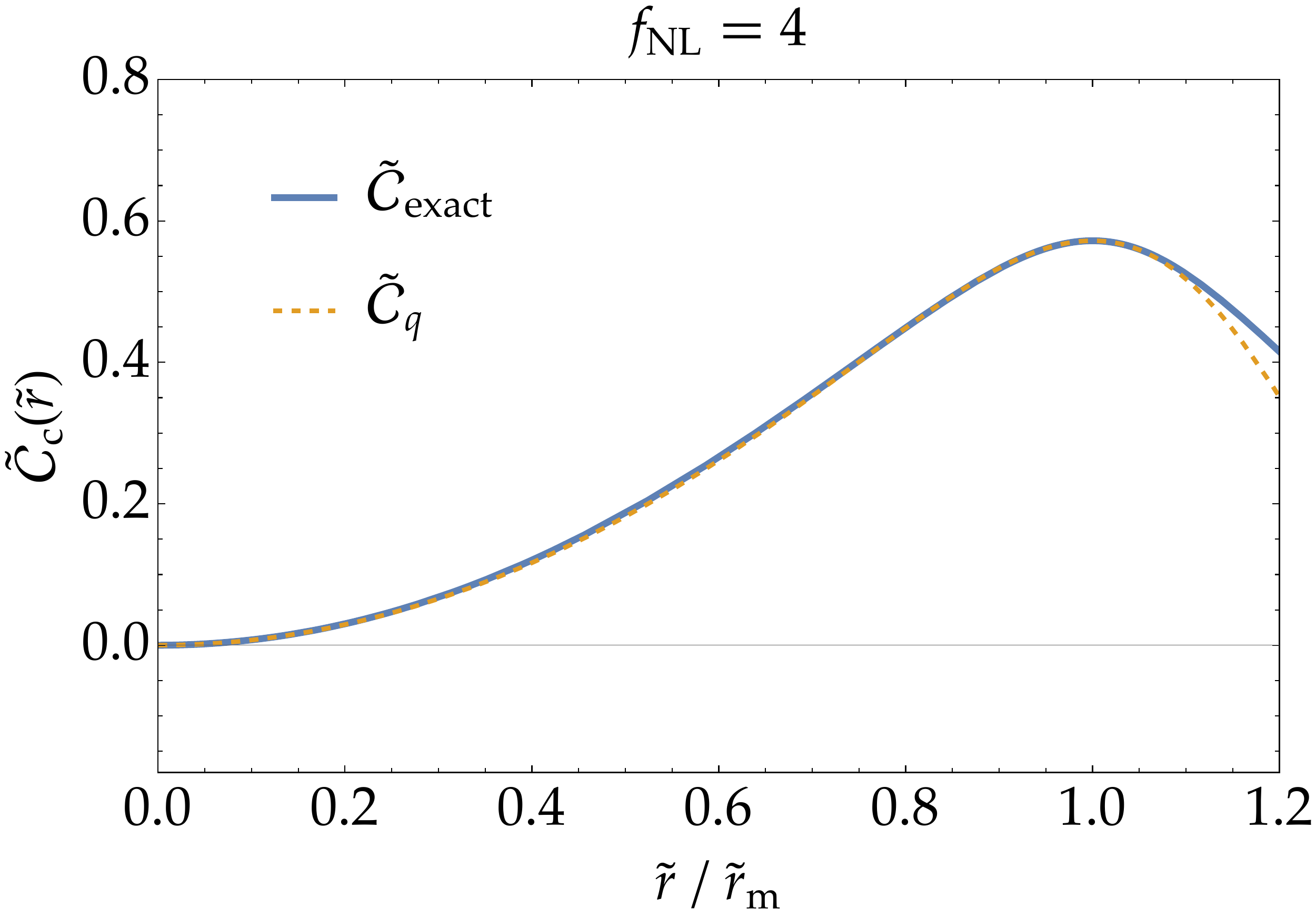}
        \end{minipage} &
        \begin{minipage}{0.45\hsize}
            \centering
            \includegraphics[width=\hsize]{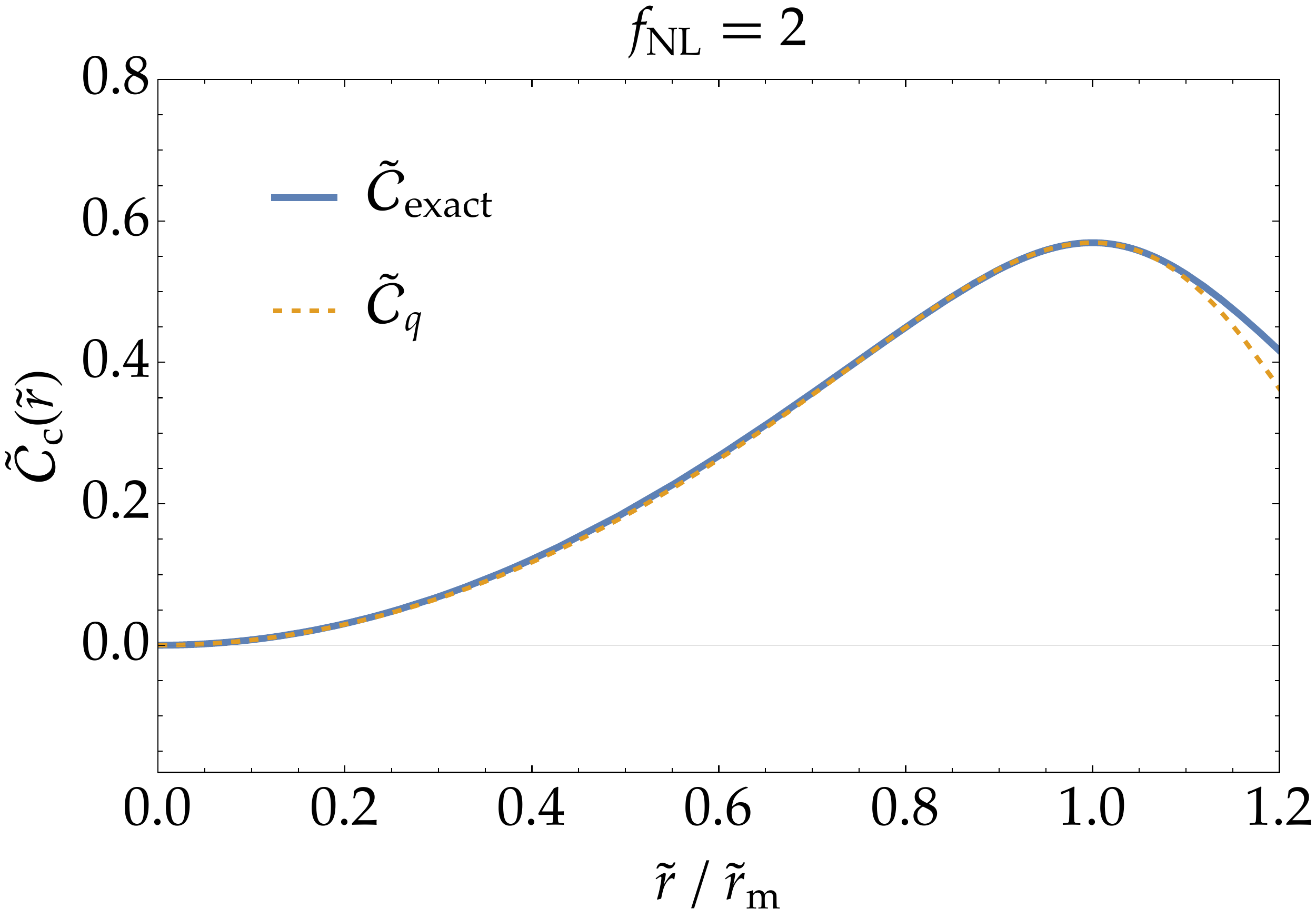}
        \end{minipage} \\\\
        \begin{minipage}{0.45\hsize}
            \centering
            \includegraphics[width=\hsize]{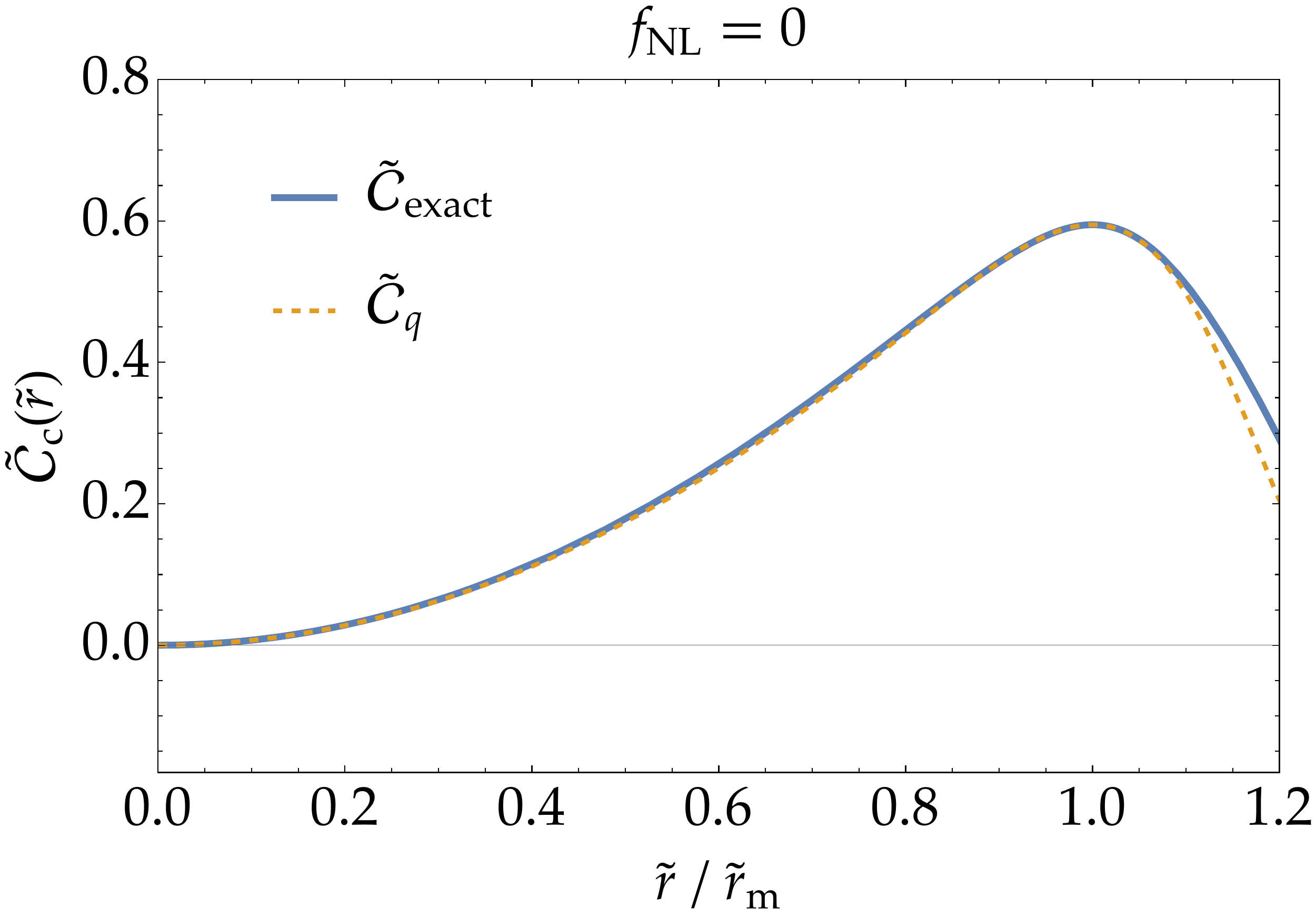}
        \end{minipage} &
        \begin{minipage}{0.45\hsize}
            \centering
            \includegraphics[width=\hsize]{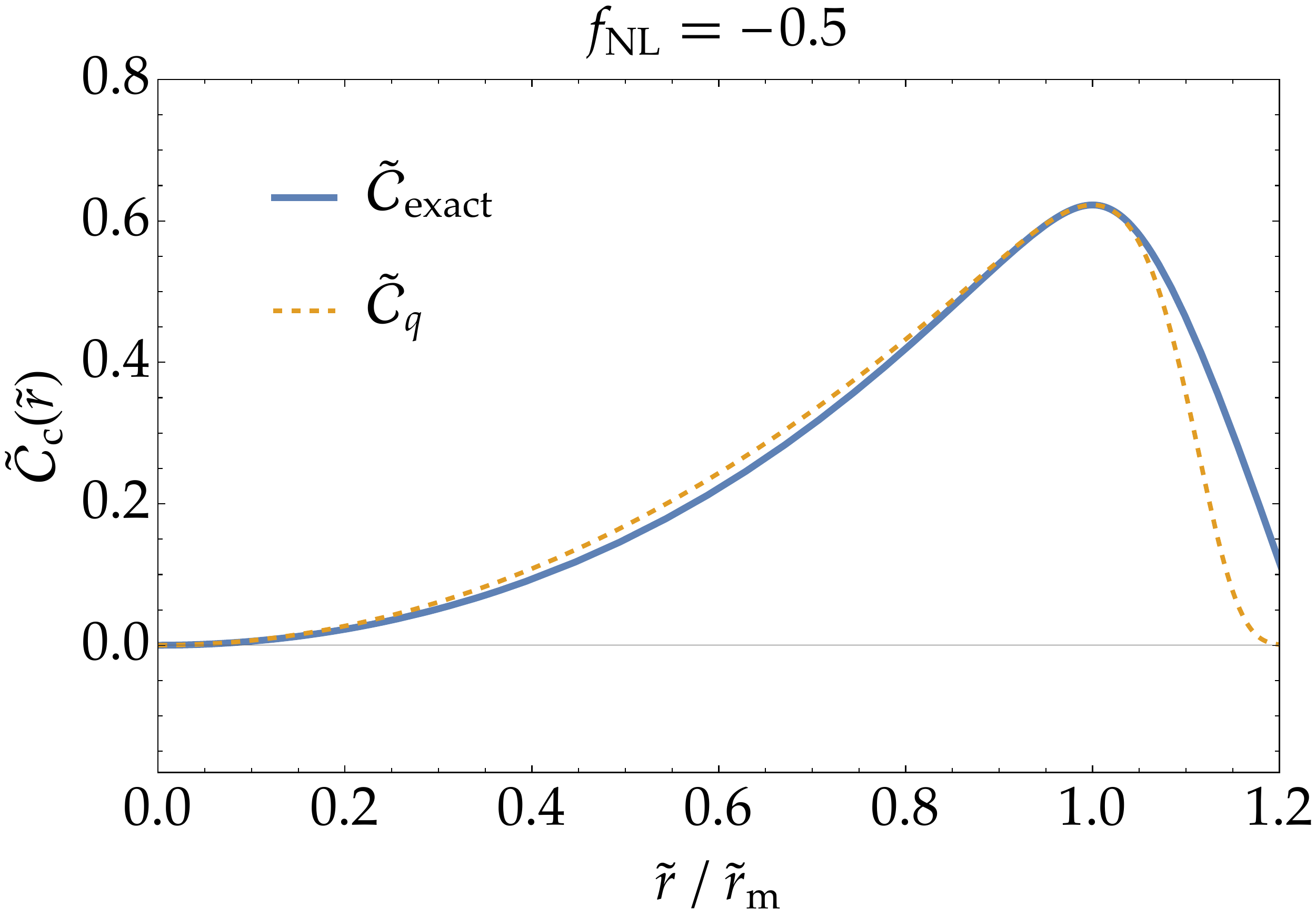}
        \end{minipage} \\\\
        \begin{minipage}{0.45\hsize}
            \centering
            \includegraphics[width=\hsize]{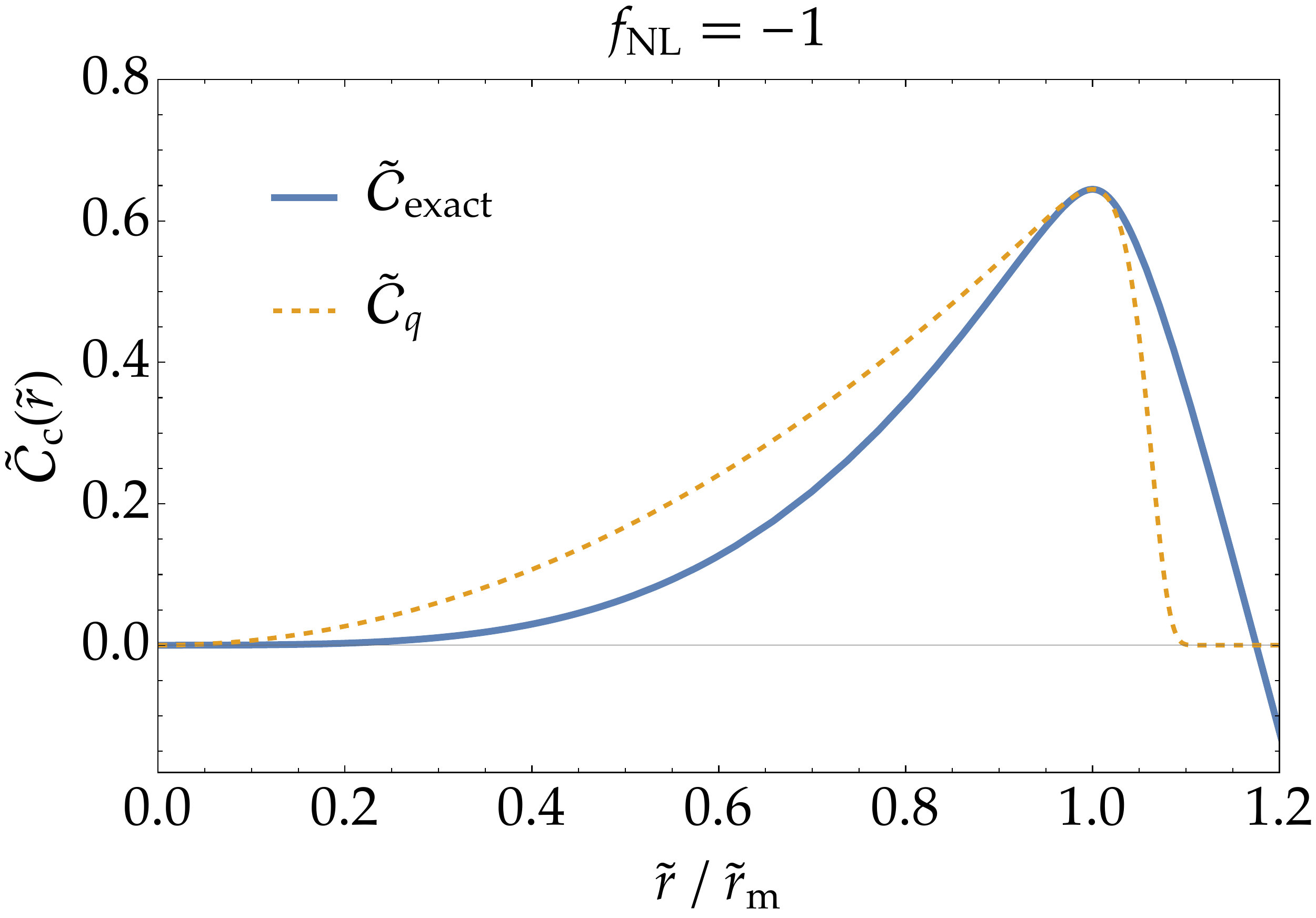}
        \end{minipage} & 
        \begin{minipage}{0.45\hsize}
            \centering
            \includegraphics[width=\hsize]{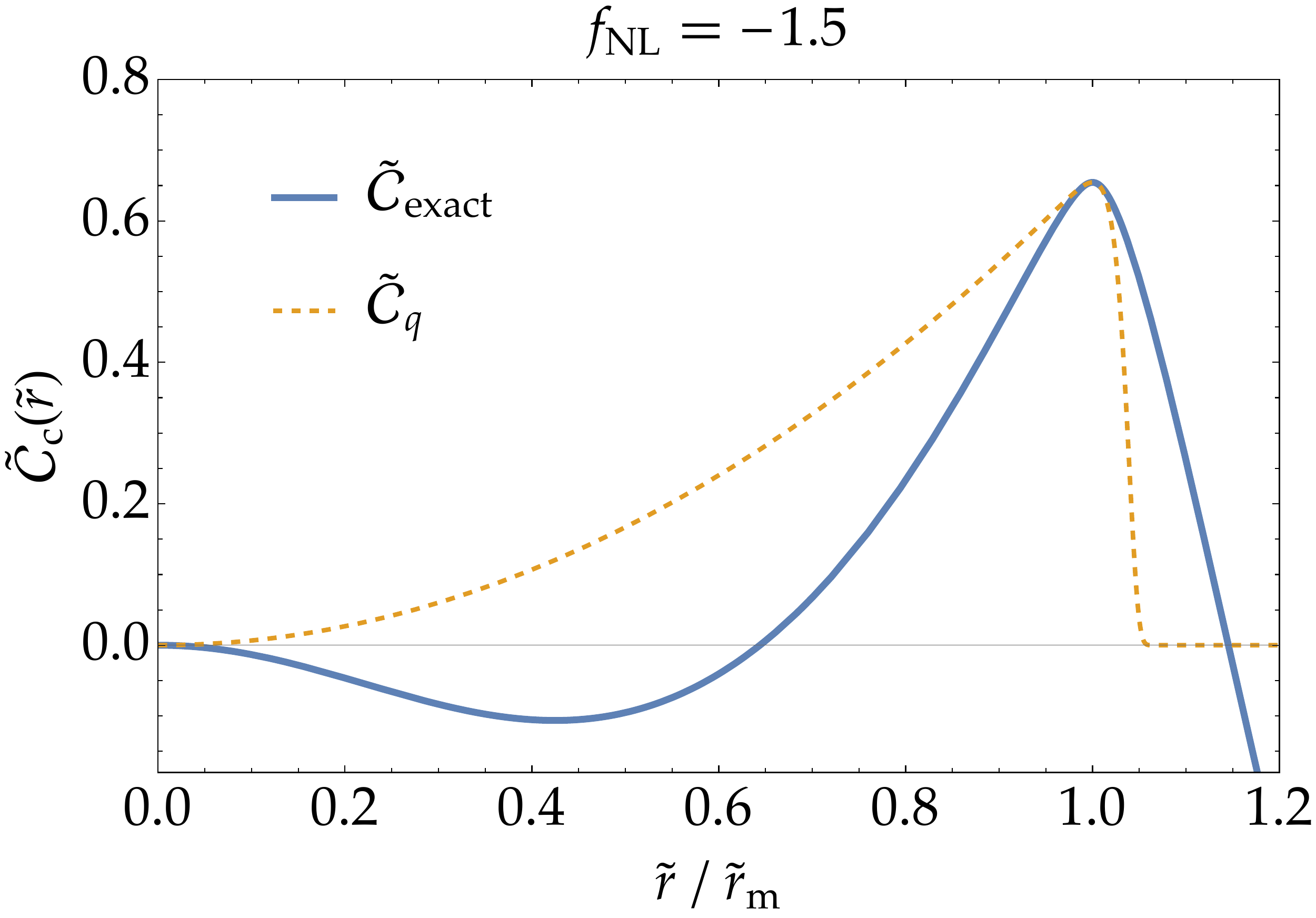}
        \end{minipage}
    \end{tabular}
    \caption{The critical compaction function with the non-Gaussian profile~\eqref{zeta_t} (blue) and the corresponding fiducial fitting~\eqref{eq: Kq} (orange-dotted, see the text for details) in terms of the comoving areal radius $\tilde{r}=r\ee^{\zeta(r)}$ normalised by the maximal radius $\tilde{r}_\um$ for several values of $\fNL$. The critical amplitude $\mu_\uc$ corresponds to the analytic threshold~\eqref{threshold_anali} (see green lines in Fig.~\ref{fig:diagram}). The fitting works well inside the maximal radius $\tilde{r}_\um$ for positive $\fNL$, while it starts to fail for negative $\fNL$. Also, the non-Gaussian profile~\eqref{zeta_t} shows a negative mass excess (corresponding to the condition $(3/5)\fNL\mu<-1/2$), which would indicate the invalidity of the profile assumption~\eqref{zeta_t} itself.
    }
    \label{fig:compaction}
\end{figure}

In Fig.~\ref{fig:compaction}, we compare 
the compaction function~\eqref{eq:compact_superhorizon} with the non-Gaussian curvature perturbations~\eqref{zeta_t}
and the corresponding fiducial fitting~\eqref{eq: Kq} for several values of $\fNL$.
The perturbation amplitude $\mu$ is set to the threshold value corresponding to $\delta_\uc(q)$~\eqref{threshold_anali}.
One finds that the fitting works well inside the maximal radius $\tilde{r}_\um$ (where $\tilde{\calC}_\uc$ is maximised) for positive $\fNL$.
For negative $\fNL$, the fitting starts to fail but we will see the analytic threshold $\delta_\uc(q)$~\eqref{threshold_anali} actually well approximate the numerical result for $\fNL\gtrsim-1$.
It does not work for $\fNL\lesssim-1$, but there one has to notice the appearance of a negative mass excess near $r=0$.
The condition of the negative mass excess appearance can be understood by checking the behaviour of $\nabla^2\zeta$ around $r=0$ because the density contrast $\delta$ is given as $\sim-\nabla^2\zeta$ at leading order in the gradient expansion (see Eq.~\eqref{eq:dens_contrast}). The non-Gaussian profile~\eqref{zeta_t} leads to $\nabla^2\zeta(r=0)=-k_*^2\mu(5+6\fNL\mu)/(5r_\um^2)$ and thus a negative mass excess appears if $(3/5)\fNL\mu<-1/2$. It can be also proved by the direct expansion of $\calC^\prime(r)$ around $r=0$ as
\bae{
    \eval{\calC^\prime(r)}_{r\to0}\approx\frac{8k_*^2(5\mu+6\fNL\mu^2)}{45r_\um^2}r+\calO(r^3).
}
However, the positive $\nabla^2\zeta(r=0)$ exactly means that the profile~\eqref{zeta_t} has no central peak
($r=0$), and for such a weird and non-well behaved profile, the $\fNL$ series expansion itself may be doubtful.

\begin{figure}
    \centering
   \includegraphics[width=0.55\hsize]{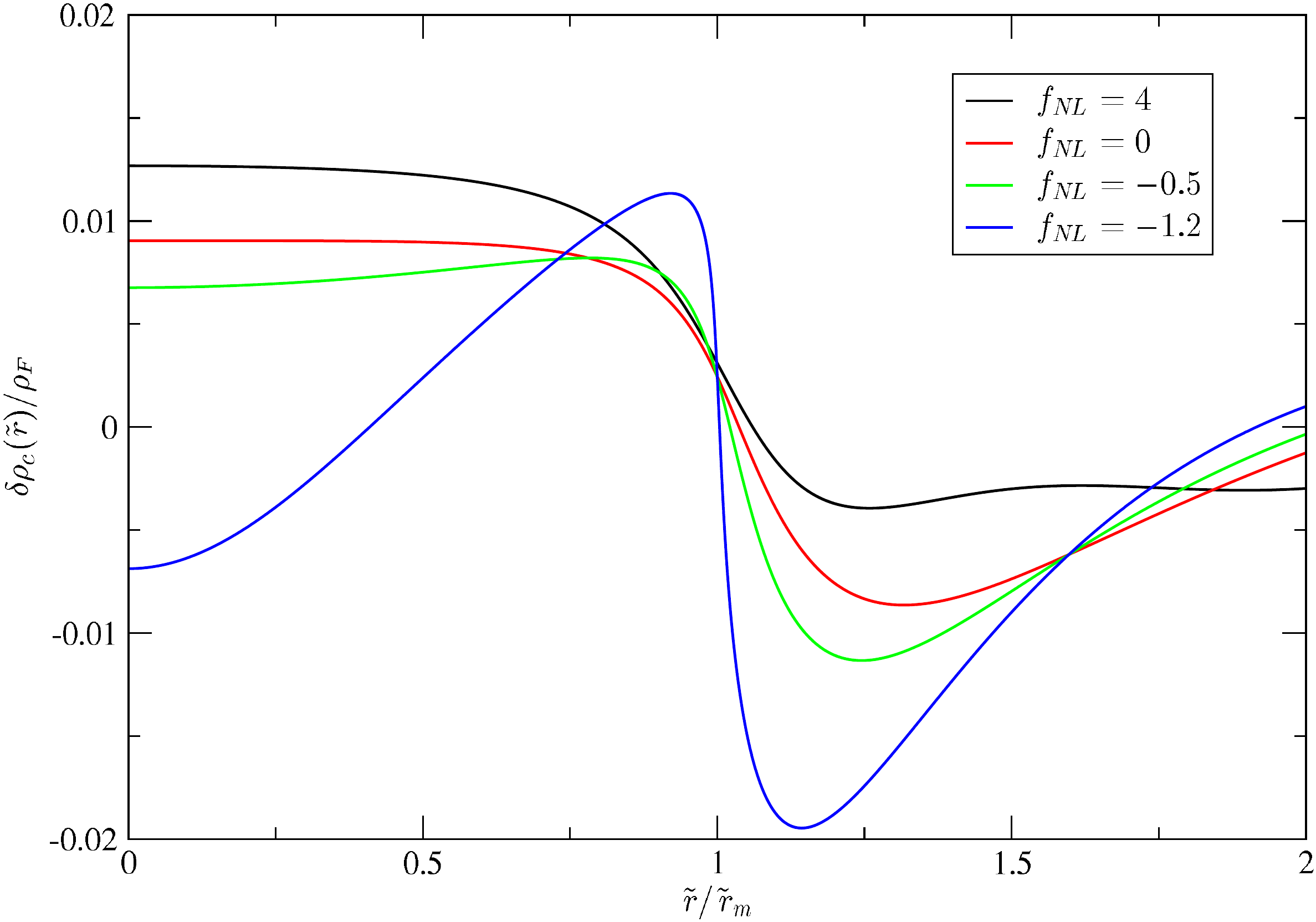}
    \caption{Critical density contrast profile for specific values of $\fNL$ in terms of $\tilde{r}$.}
    \label{fig:dens_contrast}
\end{figure}

We also show, in Fig.~\ref{fig:dens_contrast}, the density contrasts at the initial time of the simulation on superhorizon scales for different values of $\fNL$ in the coordinate $\tilde{r}$.
As peculiar properties, it shows a local minimum at the centre for a negative value of $\fNL$, and furthermore, the density contrast becomes even negative for a sufficiently negative value of $\fNL$. 
Therefore, negative values of $\fNL$ are generally counterintuitive and difficult situations, especially to produce successful numerical simulations. This behaviour is not observed for the basis profile of Eq.~\eqref{eq: Kq}.

\section{Initial conditions and set up for PBH formation}
\label{sec: Initial conditions and set up for PBH formation}

In this work, we have used the publicly available numerical code offered by Ref.~\cite{Escriva:2019nsa} to simulate numerically the formation of PBHs from the collapse of the curvature fluctuations on the FLRW universe filled by radiation fluid ($w=1/3$). The code uses Pseudospectral methods, and we refer the reader to Ref.~\cite{Escriva:2019nsa} for more details. 
Specifically, we numerically solve Misner--Sharp equations~\cite{misnersharp}, which describes the gravitational collapse of a perfect fluid with spherical symmetry. 
There the line element is generally given by, 
\bae{\label{eq:metric} 
	\dd{s^2} = -A^{2}(r,t)\dd{t^2}+B^{2}(r,t)\dd{r^2}+R^{2}(r,t)\dd{\Omega^2},
}
where $A$ is the lapse function, $R$ is the areal radius, and the radial metric $B$ is given by $B=(\partial R(r,t)/ \partial r)/\Gamma$ with the $\Gamma$ parameter defined later in Eq.~\eqref{eq: Gamma}.

The Einstein equations for the energy momentum tensor of the perfect fluid $p=w \rho$ and the metric~\eqref{eq:metric} read the following system of hyperbolic partial differential equations:
\beae{\label{eq:msequations}
	\dot{U} &= -A\left[\frac{w}{1+w}\frac{\Gamma^2}{\rho}\frac{\rho'}{R'} + \frac{M}{R^{2}}+4\pi R w \rho \right], \\ 
	\dot{R} &= A U, \\ 
	\dot{\rho} &= -A \rho (1+w) \left(2\frac{U}{R}+\frac{U'}{R'}\right), \\ 
	\dot{M} &= -4\pi A w \rho U R^{2},
}
where the lapse $A$ has been solved analytically as $A(r,t)=[\rho_\uF(t)/\rho(r,t)]^{1/4}$, which is smoothly connected to the FLRW background in $r\to\infty$.
The dot represents time derivative $\partial / \partial t$. $U$ is the Eulerian velocity defined as $U=\dot{R}/A$ and $\Gamma$ is given by 
\bae{\label{eq: Gamma}
	\Gamma = \sqrt{1+U^{2}-\frac{2M}{R}}.
}
$M$ is the so-called Misner--Sharp mass,
\bae{
	M(r,t)=\int_0^r4\pi R^2\rho\pqty{\pdv{R}{r}}\dd{r}.
}

The initial condition on the set of Eqs.~\eqref{eq:msequations} is imposed on a superHubble scale so that it is connected to the perturbed metric~\eqref{eq:metric_superhorizon}, as developed in Ref.~\cite{Polnarev:2006aa}.
There, the gradient expansion method is applied to this end.
That is, the radial dependence of the Misner--Sharp equations is expanded in the gradient parameter $\epsilon(t)$ defined by
\bae{\label{eq:epsilon}
	\epsilon(t)\equiv\frac{1}{H(t)L(t)},
}
where $H(t)$ is the Hubble factor and $L(t)\coloneqq a(t)r_\um\ee^{\zeta(r_\um)}$ is the length scale of the perturbation. 
It results in the following initial conditions~\cite{Polnarev:2006aa,Musco:2018rwt}:
\beae{\label{expansion}
	A(r,t) &= 1+\epsilon^2(t) \tilde{A}(r), \\
	R(r,t) &= a\ee^{\zeta(r)}r\qty(1+\epsilon^2(t) \tilde{R}(r)), \\ 
	U(r,t) &= H(t) R(r,t) \qty(1+\epsilon^2(t) \tilde{U}(r) ), \\ 
	\rho(r,t) &=\rho_{\uF}(t)\qty(1+\epsilon^2(t)\tilde{\rho}(r)), \\ 
	M(r,t) &= \frac{4\pi}{3}\rho_\uF(t) R(r,t)^3 \qty(1+\epsilon^2(t) \tilde{M}(r) ),\\ 
}
where 
\beae{\label{2_perturbations}
	\tilde{\rho}(r) &= -\frac{2(1+w)}{5+3w}\frac{\exp{2 \zeta(r_{\um})}}{\exp{2 \zeta(r)}}\left[\zeta''(r)+\zeta'(r)\left(\frac{2}{r}+\frac{1}{2}\zeta'(r)\right)r_{\um}^{2}\right], \\
	\tilde{U}(r) &=\frac{1}{5+3w}\frac{\exp{2 \zeta(r_{\um})}}{\exp{2 \zeta(r)}}\zeta'(r)\left[\frac{2}{r}+\zeta'(r)\right] r_{\um}^{2}, \\
	\tilde{A}(r) &= -\frac{w}{1+w} \tilde{\rho}(r), \\
	\tilde{M}(r) &= -3(1+w) \tilde{U}(r), \\
	\tilde{R}(r) &= -\frac{w}{(1+3w)(1+w)}\tilde{\rho}(r)+\frac{1}{1+3w}\tilde{U}(r). \\
}
Once the peak profile of the curvature perturbation is fixed as Eq.~\eqref{zeta_t} and the maximal radius $r_\um$ is found by Eq.~\eqref{eq:rm}, the initial conditions can be set up.
Note that the choice of different gauges should give equivalent results up to $\calO(\epsilon^2)$ as shown in Ref.~\cite{Harada:2015yda}.

The initial time of our simulations is normalised as $t_0=1$ and the background conditions are given at that time by $a(t_0)=1$, $R_H(t_0)\coloneqq1/H(t_0)=2t_0$, and $\rho_\uF(t_0)=3H^2(t_0)/8\pi$.
The characteristic time scale $t_H\coloneqq t_0(a_0r_\um/R_H(t_0))^2$ is also useful, at which time the gradient parameter reaches unity, $\epsilon(t_H)=1$.
We use three Chebyshev grids with size $N \approx 70$ (although for some cases the number of points is increased) and the boundary conditions specified in Ref.~\cite{Escriva:2019nsa}. The time step is chosen as $\dd{t}=\dd{t_{0}}(t/t_{0})^{1/2}$ with $\dd{t_{0}}=10^{-3}$. We have also ensured that for each initial configuration the epsilon parameter is less than $\epsilon(t_0) \lesssim 10^{-1}$ (this ensures that the first order in gradient expansion is enough accurate~\cite{Polnarev:2012bi}). 
In particular, it should be noted that the maximal radius $r_\um$ is equivalent to $r_*\coloneqq2.7471k^{-1}_*$ for $\fNL=0$.
Thus we choose the perturbation scale $k_*$ so that $r_*=10R_H(t_0)$ which satisfies $\epsilon(t_0)\lesssim10^{-1}$ for $\fNL=0$ and also $\fNL\neq0$ in a relevant range.

\section{Numerical results and comparison with analytical estimations}
\label{sec: Numerical results and comparison with analytical estimations}

We particularly focus our numerical simulations on the regime of non-Gaussianity where $\fNL<0$, which is the one unexplored in the literature using our approach~\cite{Kitajima:2021fpq}. It is important to mention that this regime was not numerically investigated also in Ref.~\cite{Atal:2019erb} due to the difficulty of such simulations. In this work, we have been able to do that thanks to use of multigrid domains.

\begin{figure}
    \centering
    \begin{tabular}{cc}
        \begin{minipage}{0.45\hsize}
            \centering
            \includegraphics[width=\hsize]{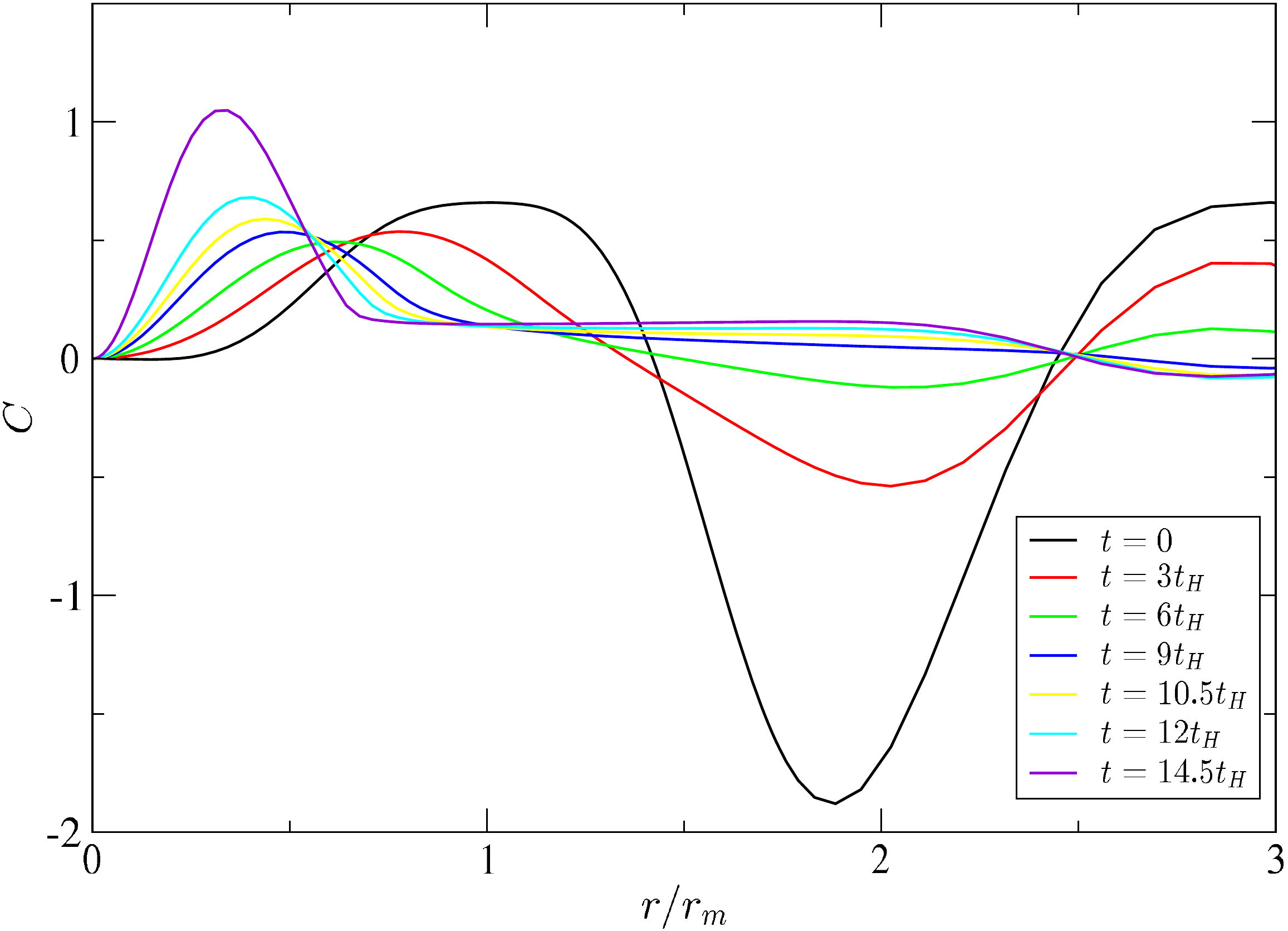}
        \end{minipage} & 
        \begin{minipage}{0.45\hsize}
            \centering
            \includegraphics[width=\hsize]{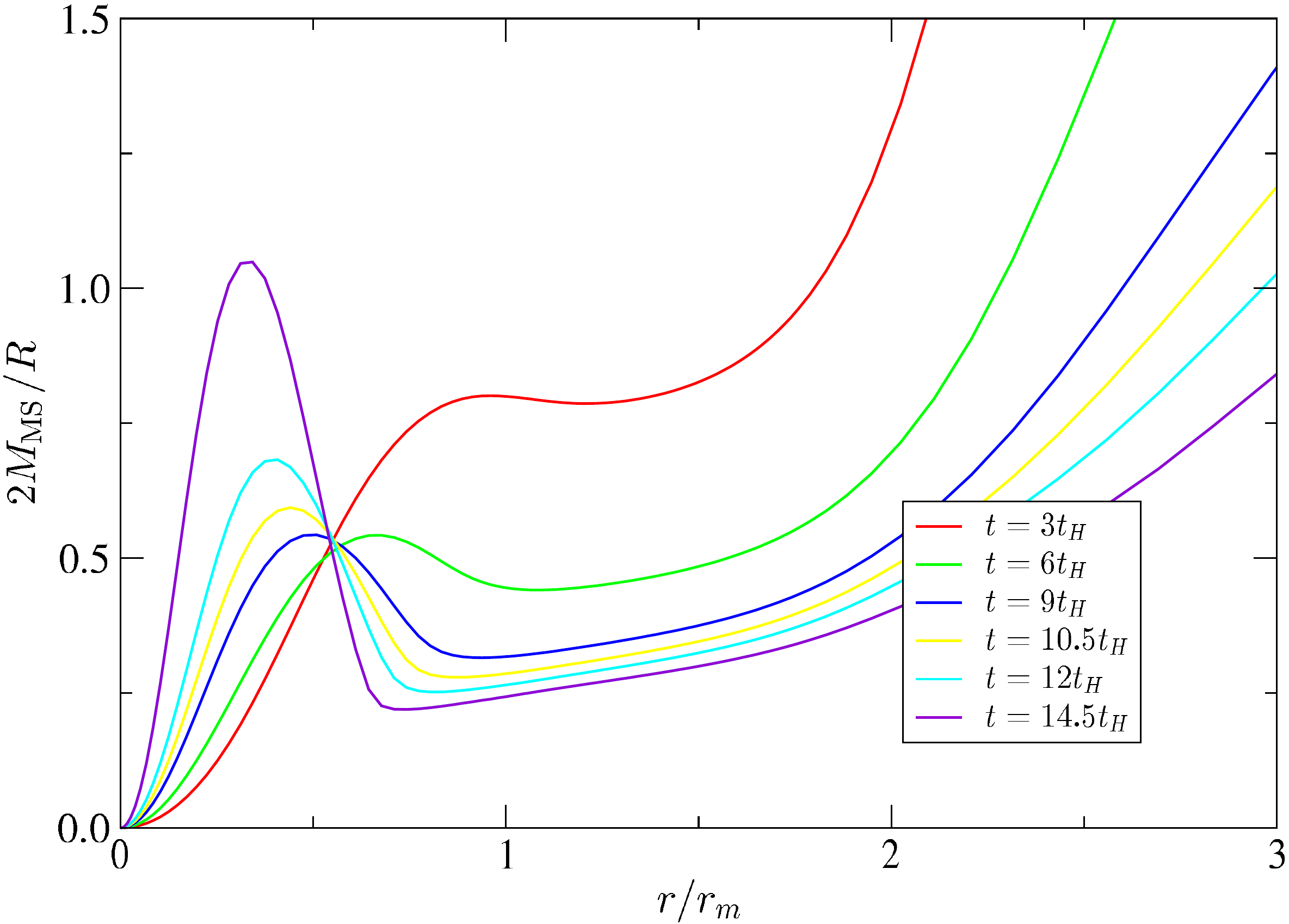}
        \end{minipage} \\\\
        \begin{minipage}{0.45\hsize}
            \centering
            \includegraphics[width=\hsize]{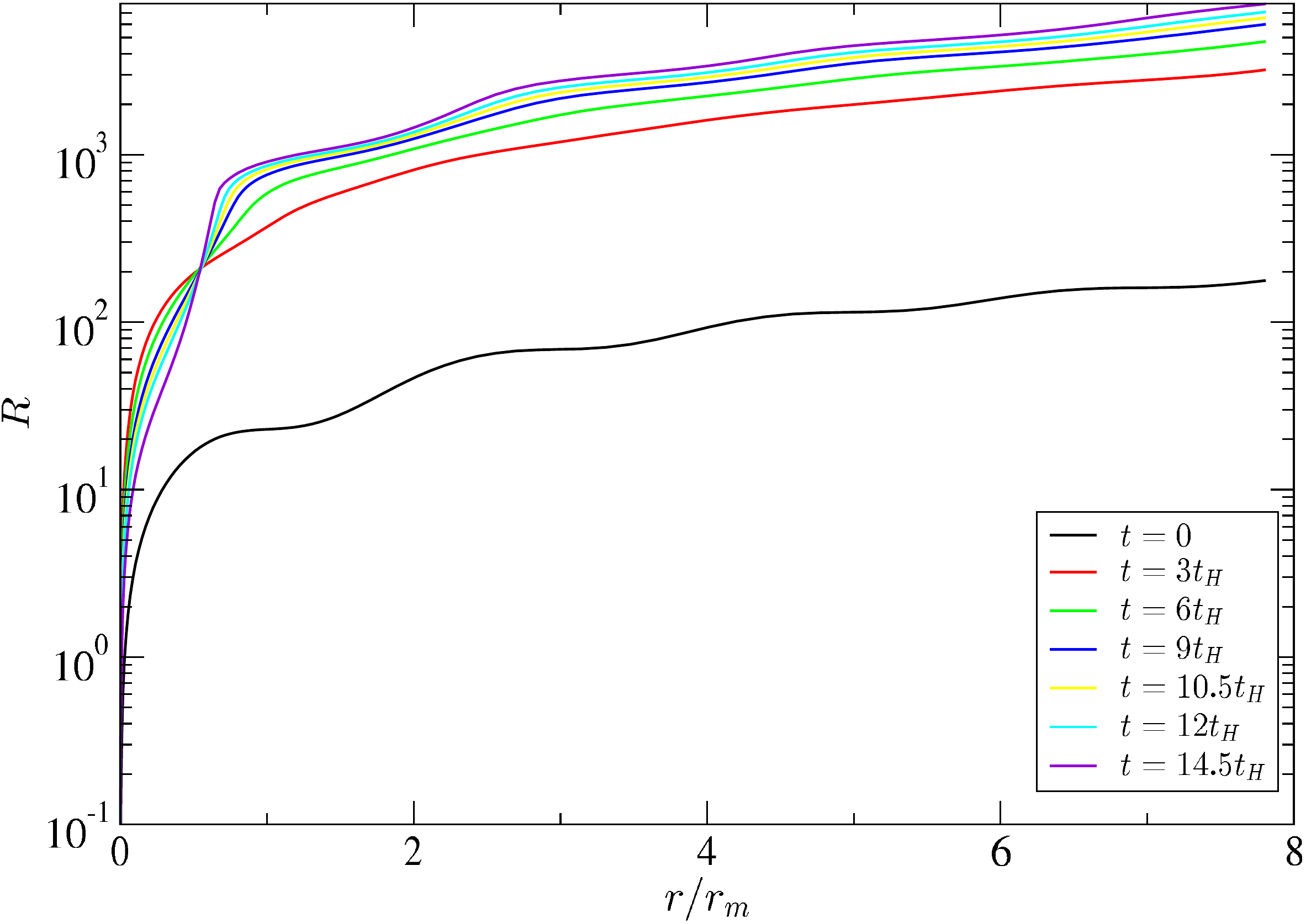}
        \end{minipage} & 
        \begin{minipage}{0.45\hsize}
            \centering
            \includegraphics[width=\hsize]{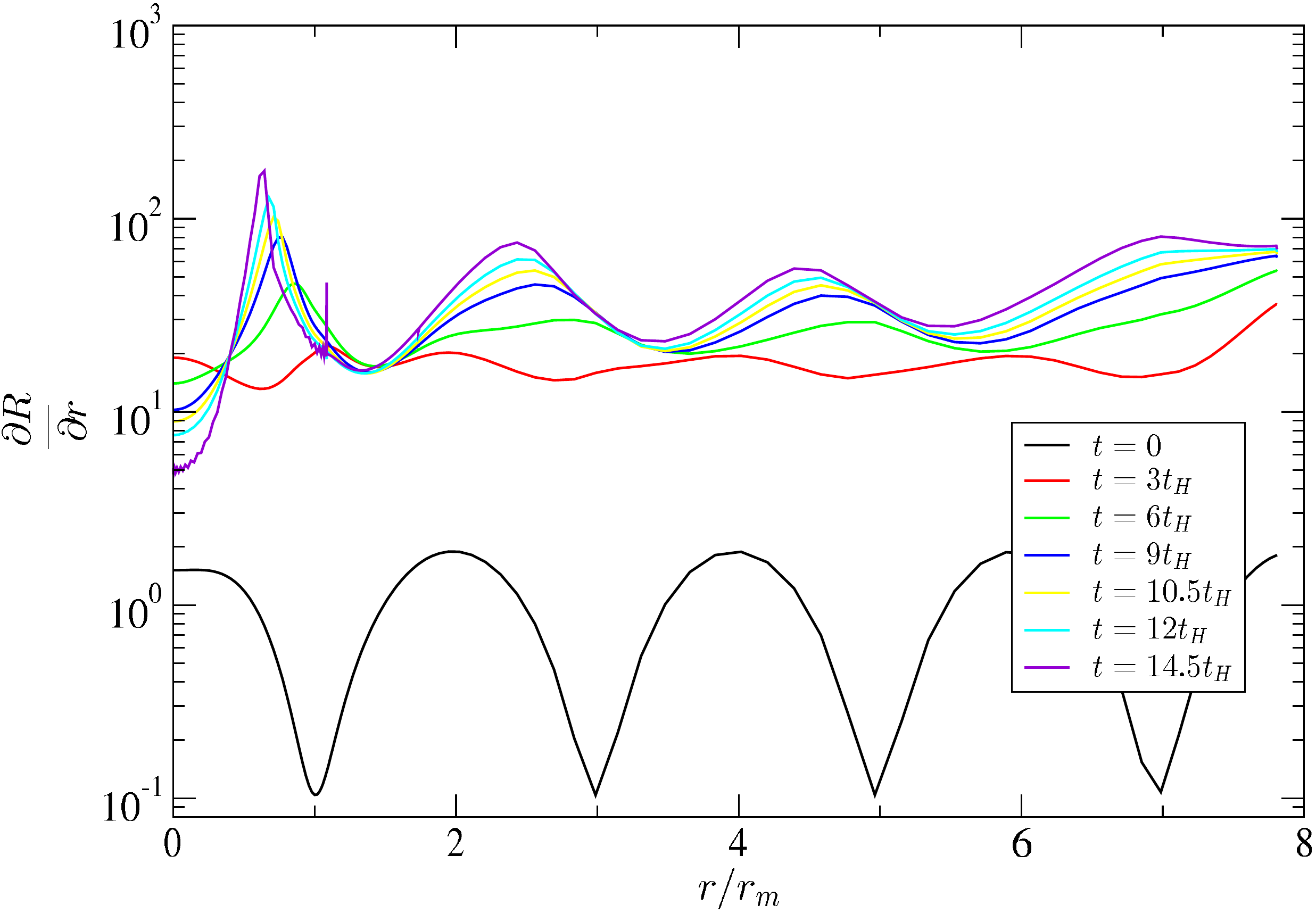}
        \end{minipage}
    \end{tabular}
    \caption{\emph{Top-Left}: the compaction function $\calC$. \emph{Top-Right}: 
    the Misner--sharp mass over the areal radius $2M_\MS/R$.  \emph{Bottom-left}: the areal radius $R$. \emph{Bottom-right}: the derivative of the areal radius $R'=\partial R / \partial r$. The different lines correspond to different times $t$, where $t_H$ is the time of horizon crossing. The initial fluctuation corresponds to $\fNL=-1$ and $\mu=0.9$ with $\mu_{\uc}=0.817$.}
    \label{fig:evolutions}
\end{figure}

In Fig.~\ref{fig:evolutions} we can see an example of the numerical evolution for a specific case with $\fNL=-1$ and $\mu=0.9$, which corresponds to a supercritical evolution ($\mu>\mu_{\uc}$) leading to BH formation. 
The compaction function $\calC$ is plotted in the top-left panel. At the initial time (on a superhorizon scale), the compaction function has several peaks outside the first one. However, the first peak's amplitude is slightly higher than the others, and therefore this is what leads to the formation of the apparent horizon. Once the perturbation crosses the horizon (it corresponds to $t>t_H$), the perturbation evolves in a fully nonlinear way. The negative/positive mass excess of the different peaks of $\calC$ (outside the first one) is smoothed out within the FLRW background whereas the mass excess of the first peak grows. For $t \approx 14 t_H$ the apparent horizon (marginally outer trapped surface) is formed when $2M_\MS/R=1$, as can be seen in the top-right panel. In the bottom panels instead, we have plotted the areal radius $R$ (left) and its derivative $R'=\partial R/\partial r$ (right). The initial conditions fulfil that $R$ is a monotonic function ($R'>0$) for type I fluctuations, and this condition holds during the whole evolution of the collapse.

\begin{figure}
    \centering
    \includegraphics[width=0.6\hsize]{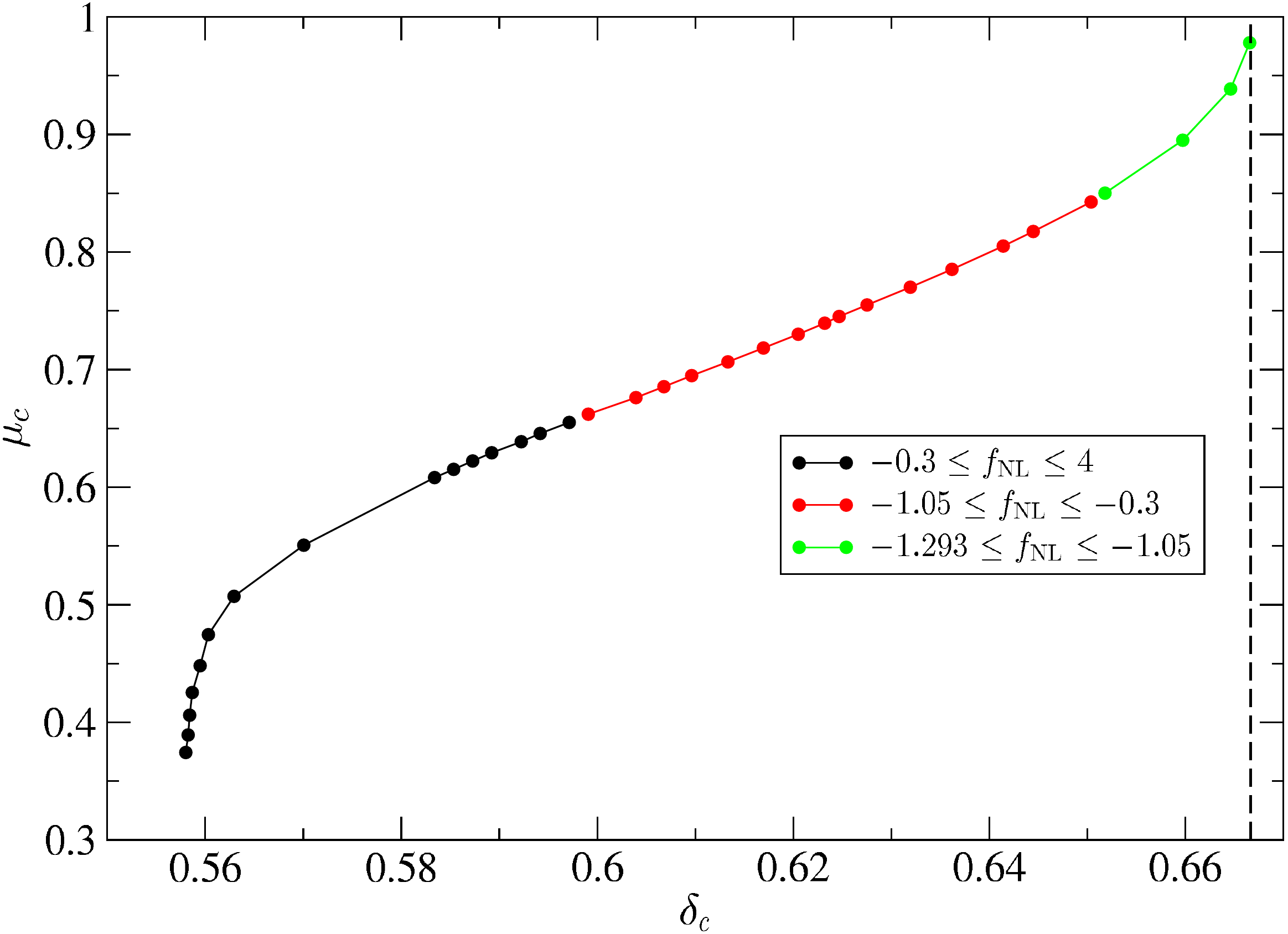}
    \caption{Critical values of the threshold $\mu_{\uc}$ in terms of $\delta_{\uc}$. The three colours represent three different regions of $\fNL$. The dashed line corresponds to the limit $\delta_{\uc,\max} = 3(1+w)/(5+3w)$ (see Eq.~\eqref{eq: delta max}).}
    \label{fig:critical_values}
\end{figure}

Using a bisection method, we have obtained the threshold values $\delta_\uc$ for different values of $\fNL$. The result can be seen in Fig.~\ref{fig:critical_values}, where the nonlinear relation between $\delta_\uc$ and $\mu_\uc$ is clear. 
It should be emphasised that we have found the existence of (type I) black hole formation for $-1.2\lesssim\fNL\lesssim-0.336$ (red and green regions in Fig.~\ref{fig:critical_values}), although Ref.~\cite{Kitajima:2021fpq} clarified that the average compaction never reaches the universal threshold $\bar{\calC}_\uc=2/5$ for $\fNL\lesssim-0.336$.
That is, it indicates the average compaction approach breaks down for negatively large non-Gaussianity $\fNL\lesssim-0.336$.

\begin{figure}
    \centering
    \begin{tabular}{cc}
        \begin{minipage}[b]{0.45\hsize}
            \centering
            \includegraphics[width=\hsize]{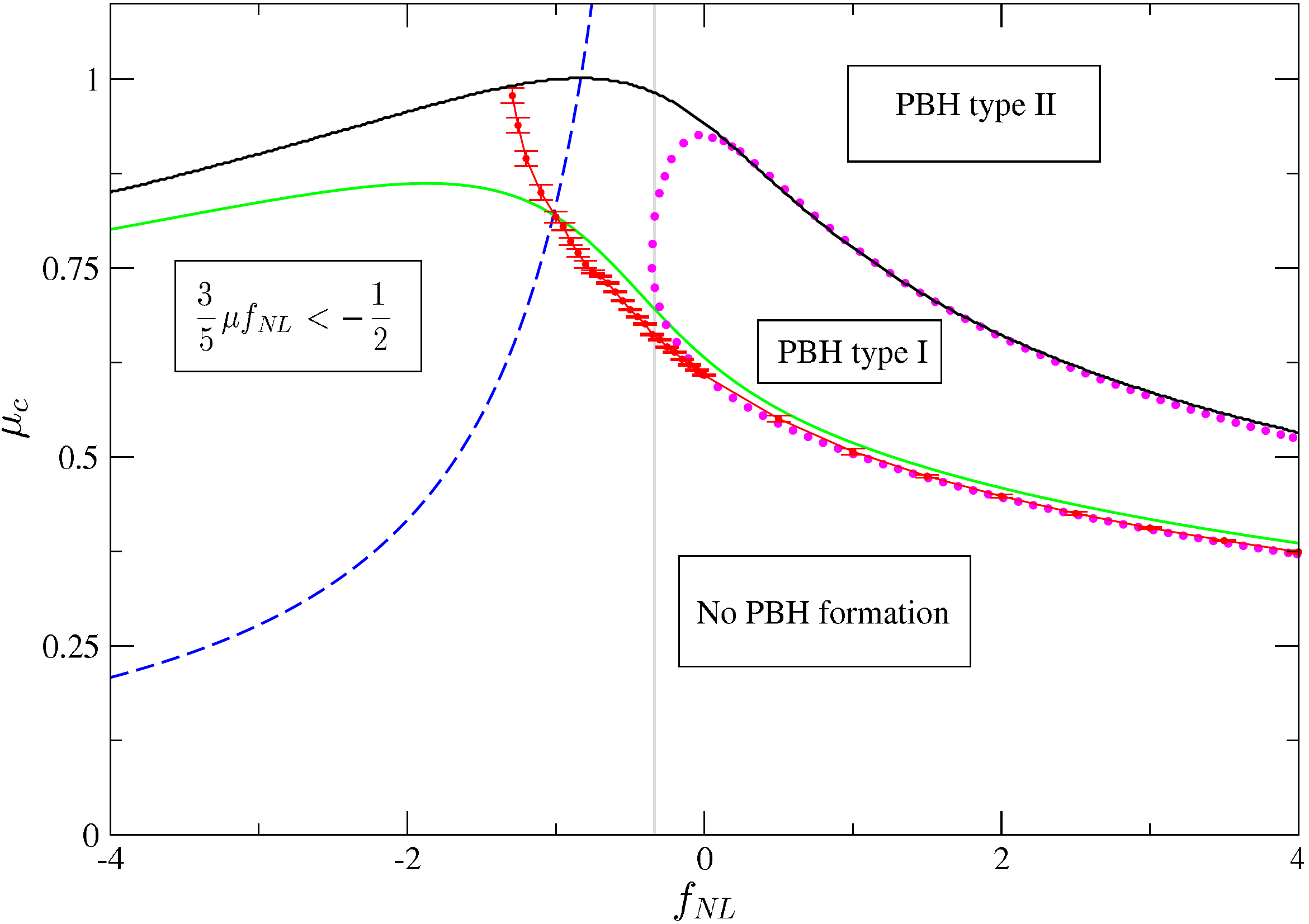}
        \end{minipage} & 
        \begin{minipage}[b]{0.45\hsize}
            \centering
            \includegraphics[width=\hsize]{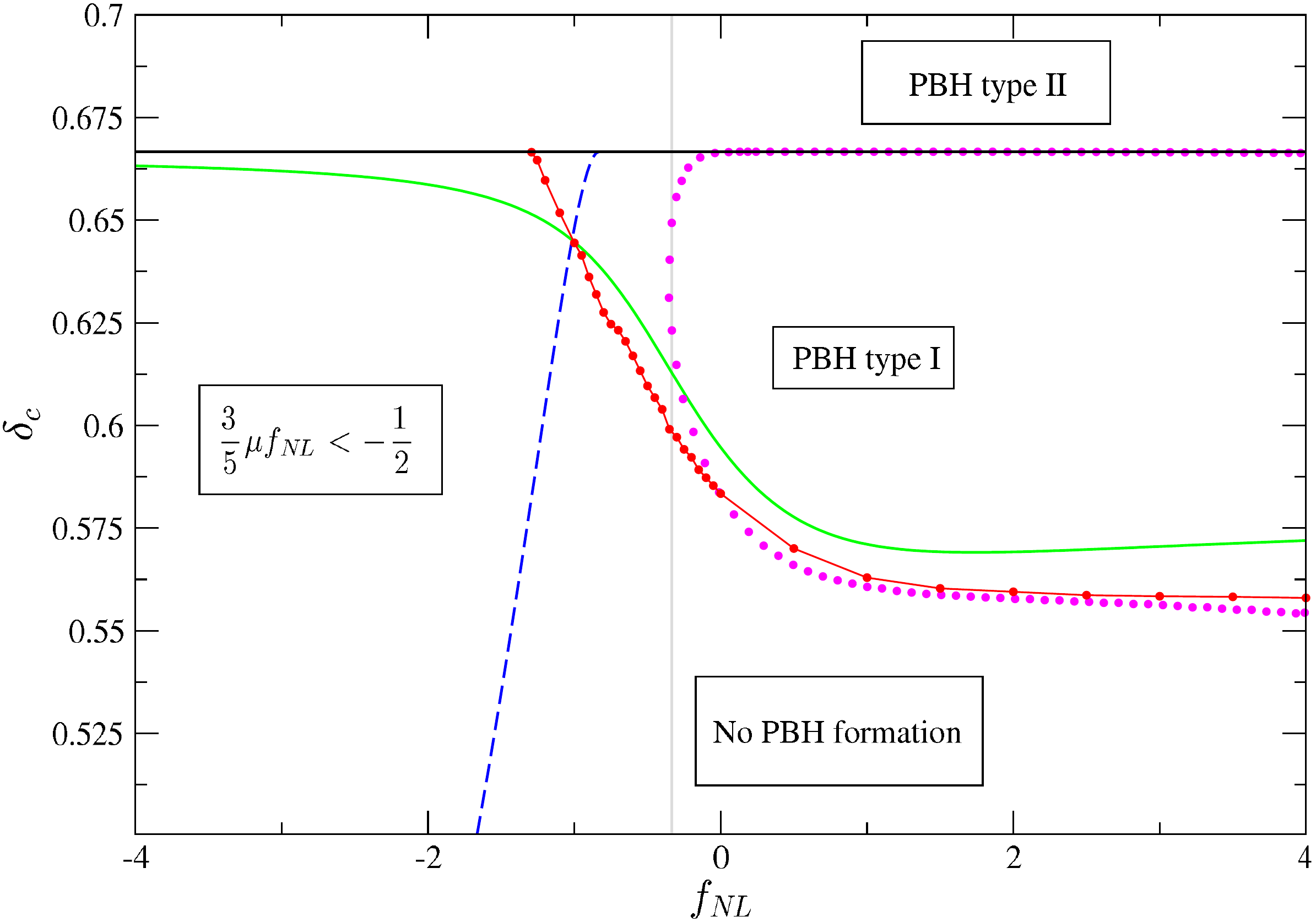}
        \end{minipage}
    \end{tabular}
    \caption{PBH diagrams in terms of $\mu_{\uc}(\fNL)$ (left) and of $\delta_{\uc}(\fNL)$ (right). The solid black lines correspond to the boundary between types I and II. Red points with error bars correspond to the numerical results of $\mu_{\uc}$ by the simulations. Dashed blue lines delimit the region where our profile assumption~\eqref{zeta_t} itself may be doubtful. The green lines show the analytical estimation of $\mu_{\uc}$ corresponding to Eq.~\eqref{threshold_anali} in the $q$-parameter approach. The magenta dotted points correspond to the threshold $\bar{\calC}_{c}= 2/5 $ in the average compaction approach. The grey vertical lines indicate the lower limit $\fNL \approx -0.336$ for the type I PBH inferred in this approach. Our numerical results reveal that the type I PBH is possible even for smaller $\fNL$.}
    \label{fig:diagram}
\end{figure}

The prediction in the average compaction approach is summarised in Fig.~5 of Ref.~\cite{Kitajima:2021fpq} as a PBH diagram.
In this work, we redraw this diagram comparing it with the numerical results, which can be found in the left panel of Fig.~\ref{fig:diagram}. 
The magenta dots correspond to the average compaction approach~\cite{Kitajima:2021fpq}, that is, the average compaction reaches the universal threshold $\bar{\calC}_\uc=2/5$.%
\footnote{The double-valued behaviour comes from the non-linear relation between the compaction function $\mathcal C$ and the curvature perturbation. In particular, in the region $R'<1$ and $r<r_{\rm m}$ for a type II perturbation, the compaction function is a decreasing function of $r$ as is easily seen by the expression $\mathcal C'$:
\beae{
\mathcal C'(r)=-\frac{4}{3}(1+r\zeta')(\zeta'+r\zeta''). 
}
} 
One can observe that the numerical results (red points with error bars) indicate the type I PBH formation even for smaller $\fNL$ than $\fNL\approx-0.336$ (grey vertical line), which is the minimum allowed $\fNL$ in the average compaction approach to indicate type I PBHs~\cite{Kitajima:2021fpq}. 
For $\fNL\lesssim-1.2$, the formation of PBHs type II could happen directly without transition to a region of PBHs type I.  
However, our profile assumption~\eqref{zeta_t} itself may be doubtful because it is in the region of $(3/5)\mu\fNL<-1/2$ as discussed at the end of section~\ref{sec:analitical}.
The critical point where the threshold $\mu_\uc$ intersects the border $(3/5)\mu\fNL=-1/2$ is found as $\fNL \approx -1.01$ and $\mu_{\uc} \approx 0.82$. 
The green lines show the analytical estimation by the $q$ parameter corresponding to Eq.~\eqref{threshold_anali}, which intriguingly gives a roughly accurate analytical description of the numerical results even for $\fNL\lesssim-0.336$ unless $(3/5)\mu\fNL<-1/2$.

The right panel of Fig.~\ref{fig:diagram} shows the corresponding behaviour to the left panel but with $\calC(r_\um)$ instead of $\mu$. One observes that the boundary that separates the two types I and II of PBH formation is given by $\delta_\um=2/3$ for any $\fNL$. 
This can be proved by taking into account that the type II perturbation is defined by the condition~\eqref{eq:conditiontype2}, i.e., it has a point such that $R^\prime<0$. The boundary between type I and II should then satisfy that there is one zero point $R^\prime=0$ and otherwise $R^\prime>0$. One finds this zero point is nothing but the maximal radius $r_\um$ by noticing that the compaction function~\eqref{eq:compact_superhorizon} can be rewritten as
\bae{\label{eq: delta max}
    \calC(r)=\frac{3(1+w)}{5+3w}\bqty{1-\pqty{\frac{R^\prime}{a\ee^{\zeta}}}^2}\leq\frac{3(1+w)}{5+3w},
}
where the equality holds if and only if $R^\prime=0$.
Accordingly, the maximal compaction (i.e., $\delta_\um$) is always given by $3(1+w)/(5+3w)=2/3$ on the boundary.
Notice that those PBHs formed for $\fNL \lesssim -0.336$ have a bigger threshold values of $\cal C$ as $\delta_\uc\gtrsim0.6$.

\begin{figure}
    \centering
    \includegraphics[width=0.6\hsize]{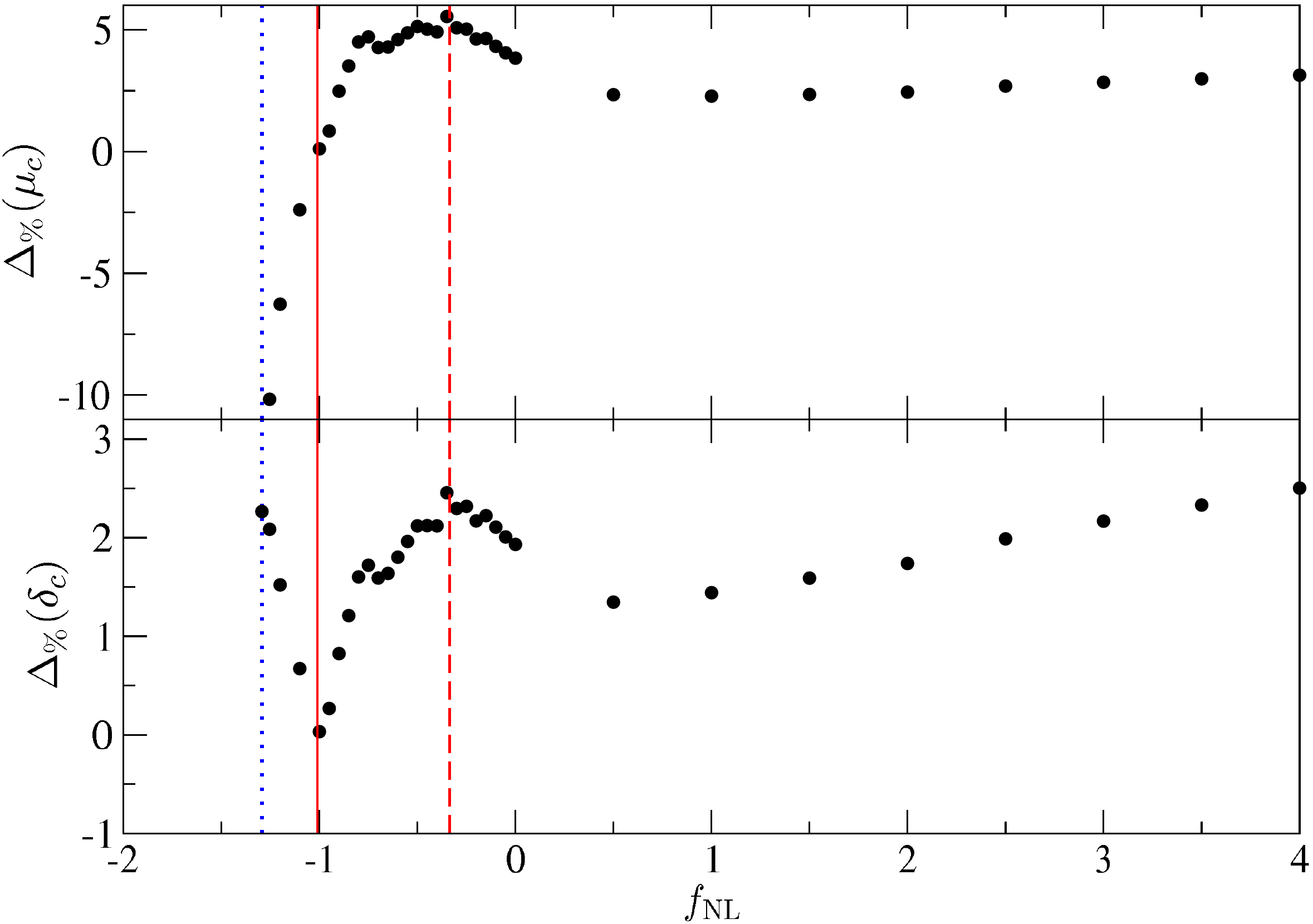}
    \caption{The percentage error in $\mu_\uc$ (top) and in $\delta_\uc$ (bottom) in terms of $\fNL$. The vertical dashed red line corresponds to $\fNL = -0.336$ (the minimum value for which was found the formation of PBHs following the average compaction approach), while the solid vertical red line to $\fNL \approx -1.01$ (the critical point where the threshold $\mu_{\uc}$ intersects the border $(3/5)\mu \fNL=-1/2$). The dotted blue line specify the boundary for type II with $\fNL \approx -1.29$.}
    \label{fig:deviation}
\end{figure}

Let us now compare the numerical results with the analytical estimations in more detail. As we have already mentioned, Ref.~\cite{Kitajima:2021fpq} adopts the averaged critical compaction function with the universal threshold $\bar{\calC}_{\uc}=2/5$ to make analytic computations.
From our numerical results in Fig.~\ref{fig:diagram}, it is however clear that  the universal criteria of $\bar{\calC}_{c}=2/5$ seems not successfully accurate for negatively large $\fNL$. 
Alternatively, in this work we have tried a different procedure to estimate the critical $\mu_{\uc}$, that is, the $q$ parameter approach using Eq.~\eqref{threshold_anali} (the green lines in Fig.~\ref{fig:diagram}) as shown in section~\ref{sec:analitical}. 
We call the $\mu_\uc$ value obtained through this approach $\mu_{\uc}^{\uA}$.
We have quantified the accuracy of the analytical estimation in comparison with the numerical results (namely $\mu_{\uc}^{\uN}$). The top panel of Fig.~\ref{fig:deviation} shows the deviation between $\mu_{\uc}^{\uN}$ and $\mu_{\uc}^{\uA}$ values with
\bae{
	\Delta_\%(\mu) = 100 \times \frac{\mu_{\uc}^{\uA}-\mu_{\uc}^{\uN}}{\mu_{\uc}^{\uN}}.
}
The same is applied to $\delta_{\uc}^{\uN}$ and $\delta_{\uc}^{\uA}$ in the bottom panel. The accuracy obtained with $\delta_{\uc}^{\uA}$ is within the range of validity found in Ref.~\cite{Escriva:2019phb}, i.e., $\lesssim\calO(2\%)$, even for the negative $\fNL$. Nevertheless, due to the nonlinear relation between $\delta_{\uc}$ and $\mu_{\uc}$, the deviation in $\mu_\uc$ is larger. The approximation in $\mu$ is roughly accurate until $\fNL \lesssim -1$ with a deviation of $\calO(4.5\%)$ for small negative $\fNL$ and of 
$\calO(2.5\%)$ for $\fNL>0$. It clearly fails for $\fNL < -1$, where our profile assumption~\eqref{zeta_t} itself may be doubtful, though.

\begin{figure}
    \centering
    \includegraphics[width=0.6\hsize]{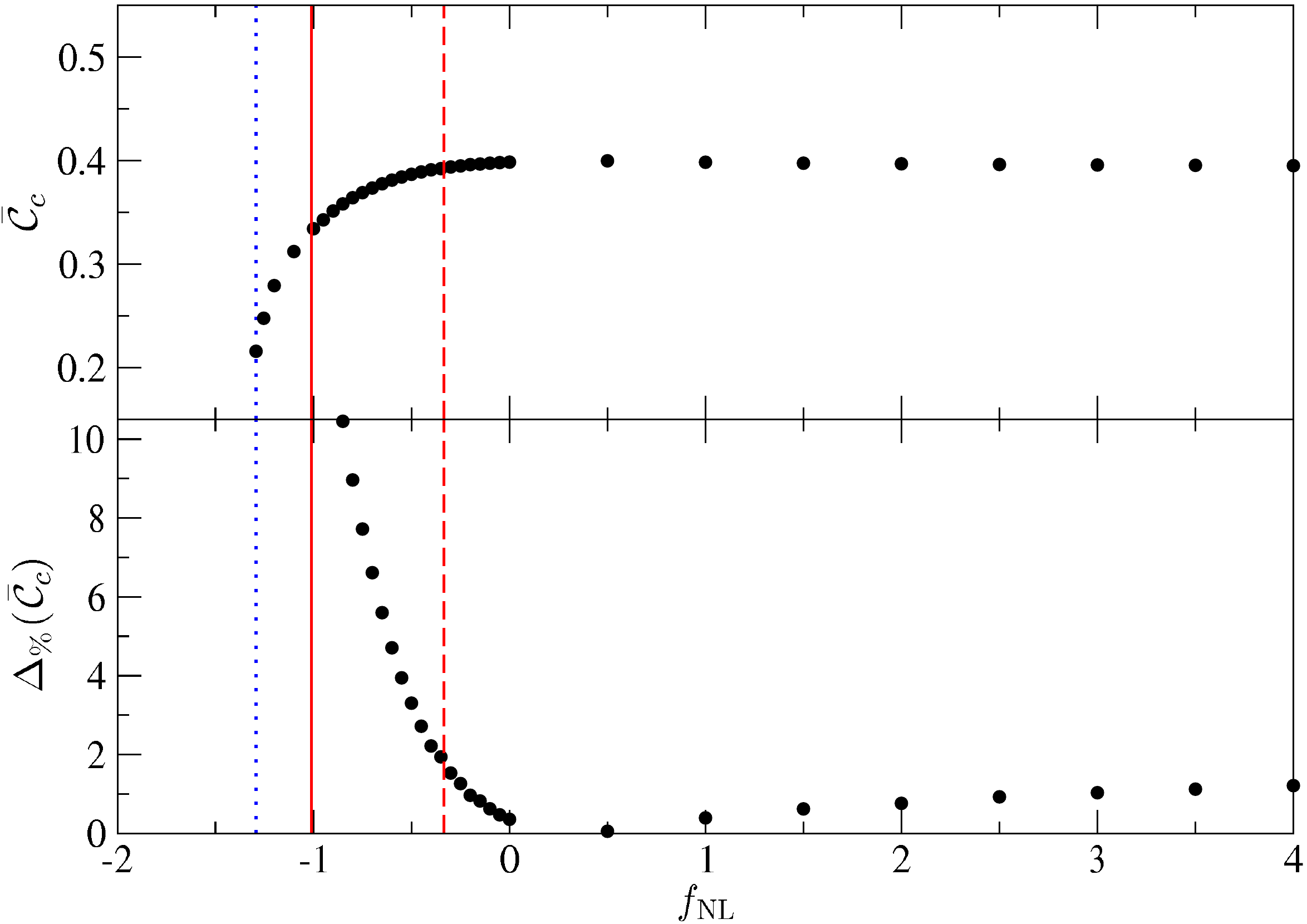}
    \caption{Numerical values of $\bar{\calC}_{\uc}$ (top) and their percentage errors to $2/5$ (bottom) in terms of $\fNL$. The vertical dashed red line corresponds to $\fNL = -0.336$ (the minimum value for which was found the formation of PBHs following the average compaction approach), while the solid vertical red line to $\fNL \approx -1.01$ (the critical point where the threshold $\mu_{\uc}$ intersects the border $(3/5)\mu \fNL=-1/2$). The dotted blue line specify the boundary for type II with $\fNL \approx -1.29$.}
    \label{fig:deviation_average}
\end{figure}

We have also checked the deviation of the critical averaged compaction function $\bar{\calC}_\uc$ in the simulations from the universal criteria $2/5$ expected in Ref.~\cite{Escriva:2019phb}. The error is defined by
\bae{
	\Delta_\%(\bar{\calC}_{\uc}) = 100 \times\frac{(2/5)-\bar{\calC}_{\uc}}{(2/5)}.
}
The result can be found in Fig.~\ref{fig:deviation_average}. The top panel shows the numerical result of $\bar{\calC}_{\uc}$ corresponding to $\mu_{\uc}^\uN$, and the relative deviation is shown in the bottom panel. 
The average compaction starts to deviate beyond $2\%$ for $\fNL \lesssim -0.336$ and the error increases substantially for smaller $\fNL$. In view of these results, it is clear that the procedure of a universal critical average compaction function seems to fail for some specific and non-well behaved curvature profiles. Instead, the procedure with the analytical estimation $\delta_{\uc}(q)$ still seems to work correctly for our purposes. The determination of $\mu_{\uc}^{\uA}$ using Eq.~\eqref{threshold_anali} is more robust than the averaged compaction~\eqref{eq: barCm}. 
We note that $\delta_\uc(q)$ is originally derived from the assumption $\bar{\calC}_\uc=2/5$~\cite{Escriva:2019nsa}. However, $\delta_\uc(q)$ is found to work beyond the average compaction assumption and thus we can consider it as a ``fitting formula" independent of the particular value of $\bar{\calC_{c}}$ chosen.

\section{PBH mass function}\label{sec:mass}

Having clarified the success of the analytical criterion of the PBH formation with numerical simulations even for negative $\fNL$, let us show some example PBH mass functions in this section.
We employ the analytic threshold $\mu_\uc$ via the $q$ parameter (the green lines in Fig.~\ref{fig:diagram}) and follow the peak theory summarised in Ref.~\cite{Kitajima:2021fpq}.

Let us first review the so-called critical behaviour for the PBH mass. That is, given the perturbation amplitude $\mu$ for an overdense region, the resultant PBH mass $M$ is assumed to follow the scaling relation~\cite{choptuik,coleman,renormalizationcriticalcollapse,Niemeyer1,Niemeyer2,hawke2002,musco2009},
\bae{
    M=K(\mu-\mu_\uc)^\gamma M_H,
}
with an order-unity parameter $K$, the universal power $\gamma\simeq0.36$, and the horizon mass $M_H$ at the reentry of the perturbation $\epsilon(t)=1/H(t)L(t)=1$.
The coefficient $K$ slightly depends on the peak profile but it has not been precisely clarified yet (see, e.g., Refs.~\cite{Escriva:2019nsa,Escriva:2021pmf} for relevant works). We hence simply adopt $K\simeq1$ in this paper.
In the case of the monochromatic power~\eqref{eq: monochromatic P}, it is helpful to define the mass $M_{k_*}$ by the horizon mass at the horizon reentry of the scale $k_*$ in the background universe. It can be obtained as (see, e.g., Ref.~\cite{Tada:2019amh})
\bae{
    M_{k_*}\simeq10^{20}\pqty{\frac{g_*}{106.75}}^{-1/6}\pqty{\frac{k_*}{\SI{1.56e13}{Mpc^{-1}}}}^{-2}\,\si{g},
}
where $g_*$ is the effective degrees of freedom for the energy density of the cosmic fluid at the horizon reentry and we assume that it is almost equivalent to those for entropy density.
As the perturbation scale is given by $L(t)=a(t)r_\um\ee^{\zeta(r_\um)}$, the PBH mass can be rewritten as
\bae{\label{eq: M in Mk*}
    M=K(\mu-\mu_\uc)^\gamma(k_*r_\um)^2\ee^{2\zeta(r_\um)}M_{k_*}.
}   
In Fig.~\ref{fig: M vs mu}, we show the PBH mass as a function of the perturbation amplitude $\mu$ for $\fNL=-1$, $0$, and $1$.

\bfe{width=0.6\hsize}{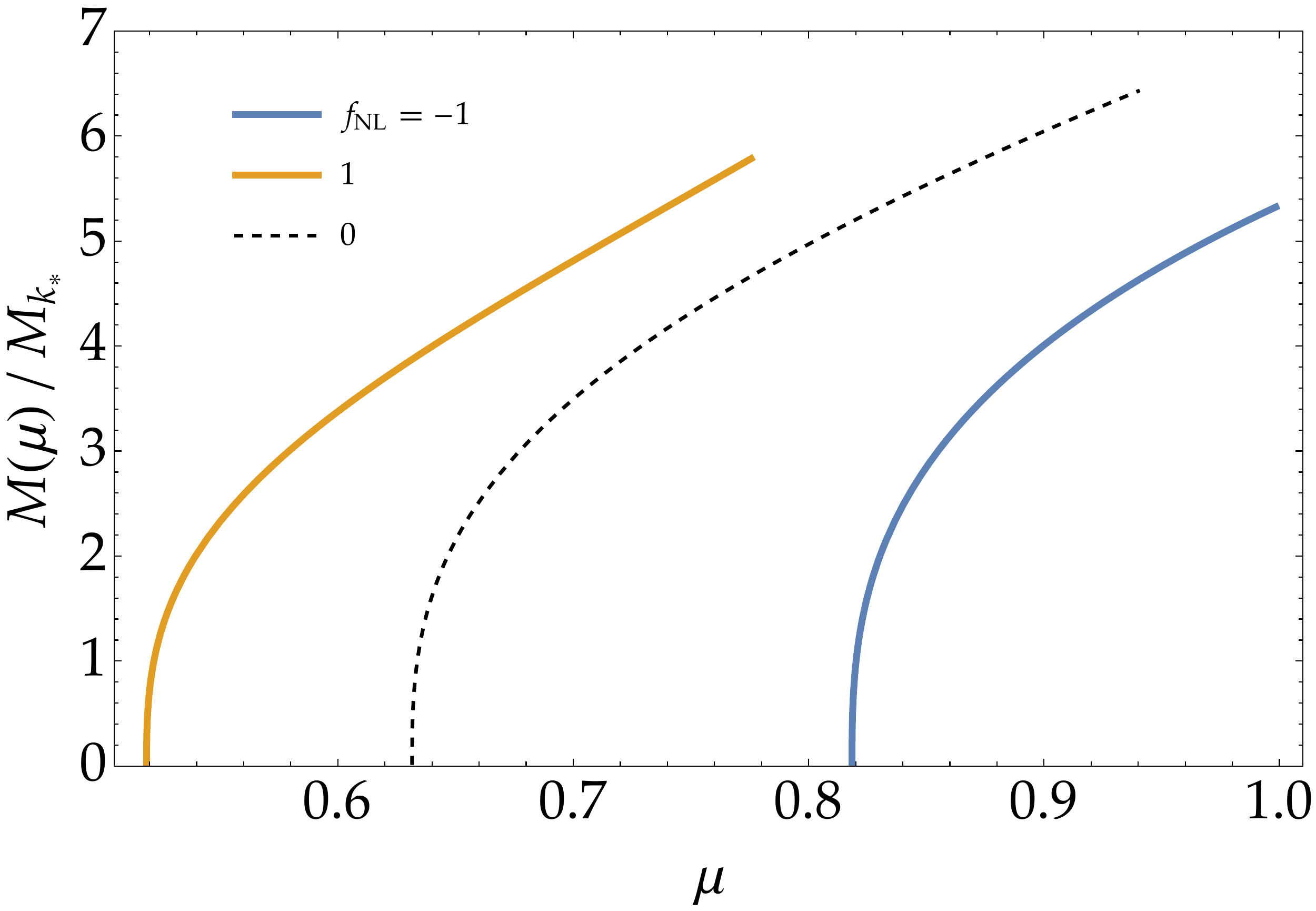}{The PBH mass $M$ in the unit of $M_{k_*}$ given by the scaling relation~\eqref{eq: M in Mk*} as a function of the perturbation amplitude $\mu$ for $\fNL=-1$ (blue), $0$ (black dashed), and $1$ (orange).}{fig: M vs mu}

\begin{figure}
	\centering
	\begin{tabular}{c}
		\begin{minipage}[b]{0.5\hsize}
			\centering
			\includegraphics[width=0.95\hsize]{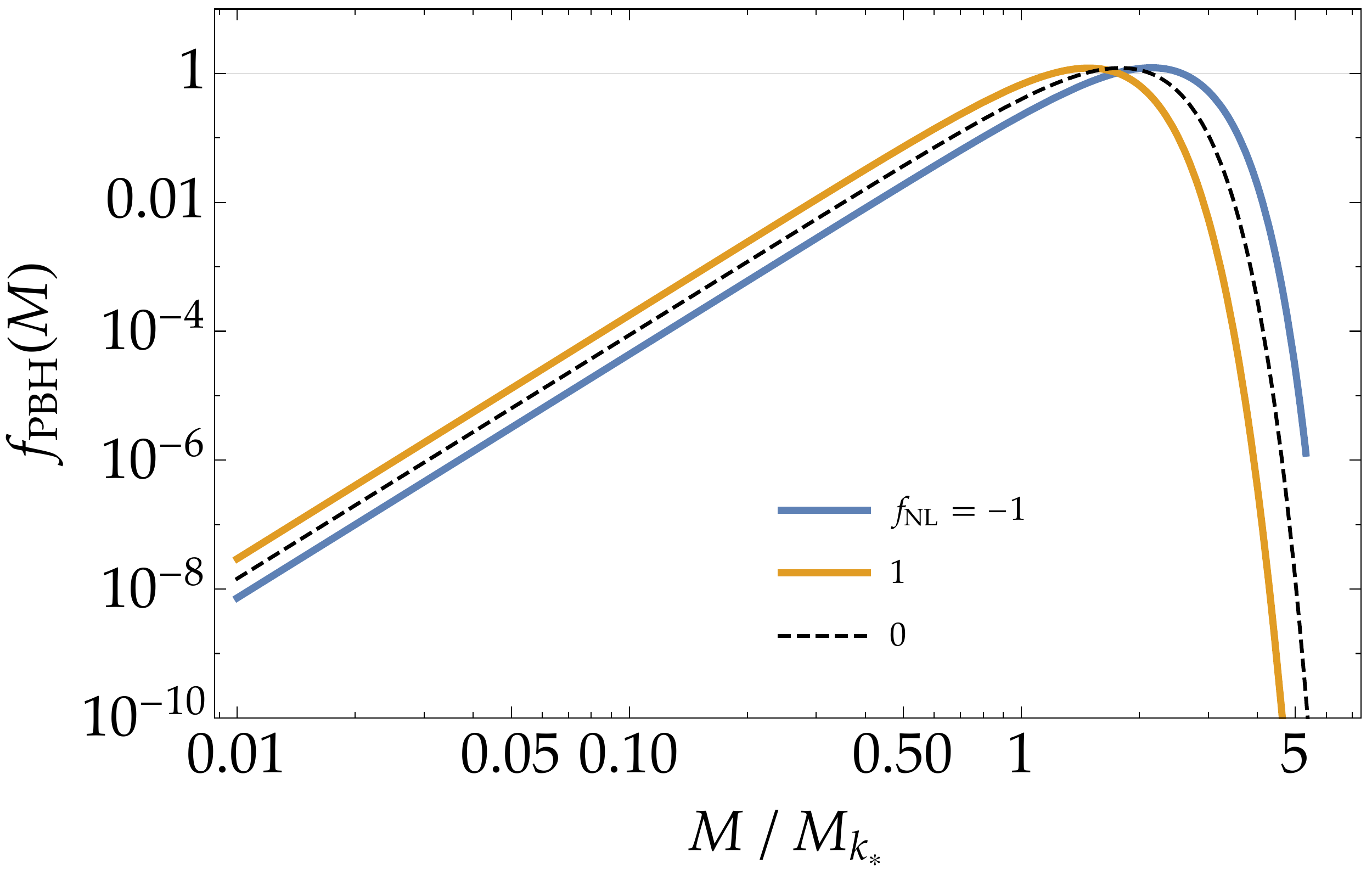}
		\end{minipage}
		\begin{minipage}[b]{0.5\hsize}
			\centering
			\includegraphics[width=0.95\hsize]{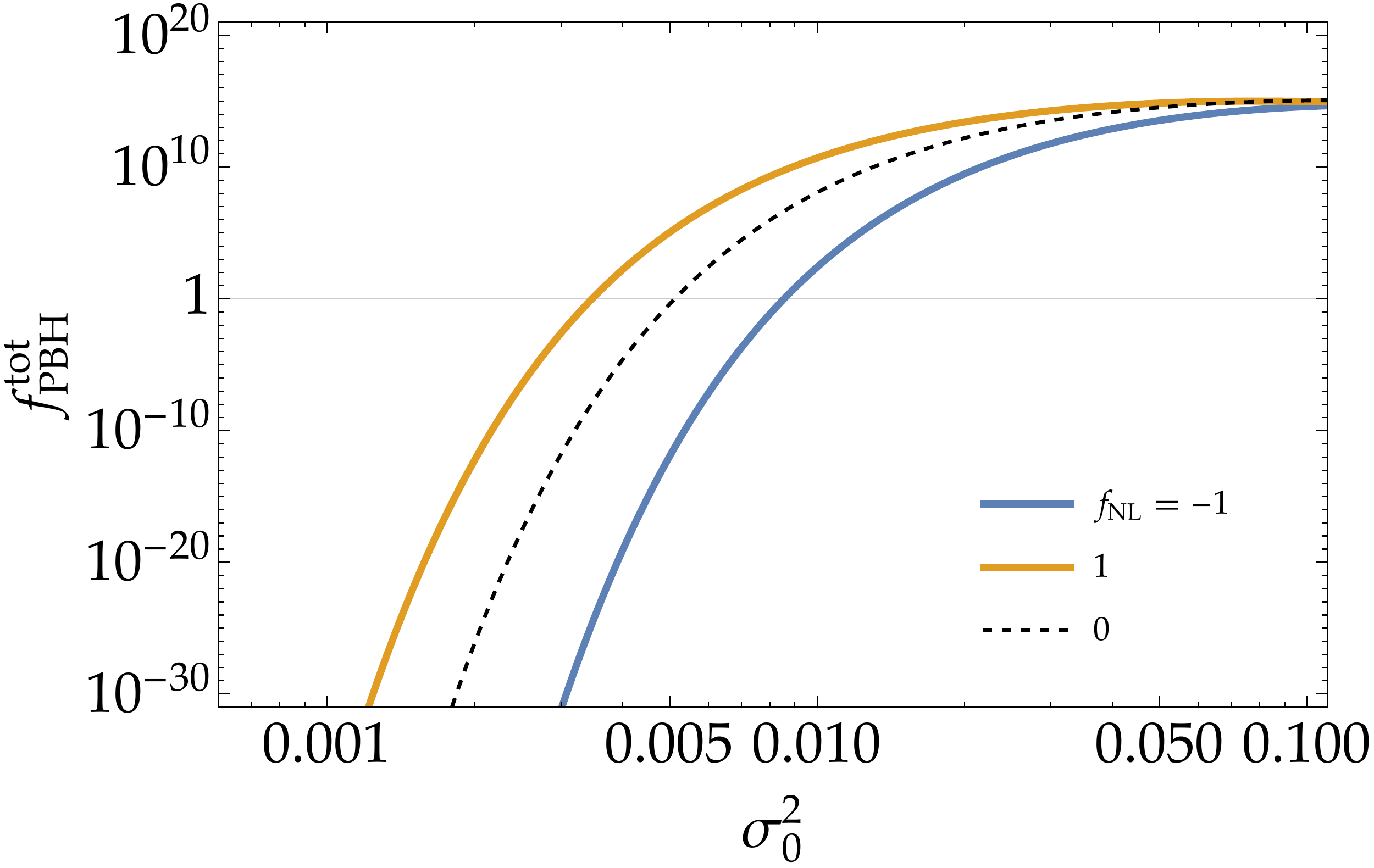}
		\end{minipage}
	\end{tabular}
	\caption{\emph{Left}: the PBH mass spectra~\eqref{eq: fPBH} for $\fNL=-1$ (blue), $0$ (black dashed), and $1$ (orange) with $k_*=\SI{1.56e13}{Mpc^{-1}}$. The variance $\sigma_0^2$ is tuned so that dark matters are fully comprised of PBHs for each $\fNL$. \emph{Right}: the total PBH abundance $f_\PBH^\tot$ as a function of $\sigma_0^2$ with the same colour code to the left panel.}
	\label{fig: fPBH}
\end{figure}

The peak number density with the amplitude $\mu$ can be statistically obtained in the peak theory. As $\mu$ is related to the PBH mass through the critical behaviour~\eqref{eq: M in Mk*}, such a peak number density can be recast into the current PBH energy density within the mass range $[M,M\ee^{\dd{\ln M}}]$.
Normalised by the current dark matter energy density, it can be calculated as (see Ref.~\cite{Kitajima:2021fpq} for the detailed derivation)
\bme{\label{eq: fPBH}
    f_\PBH(M)\dd{\ln M}=\frac{\rho_\PBH(M)}{\rho_\DM}\dd{\ln M} \\
    =\pqty{\frac{\Omega_\DM h^2}{0.12}}^{-1}\pqty{\frac{M}{10^{20}\,\si{g}}}\pqty{\frac{k_*}{\SI{1.56e13}{Mpc^{-1}}}}^3\pqty{\frac{\abs{\dv{\ln M}{\mu}}^{-1}f\pqty{\frac{\mu(M)}{\sigma_0}}P_\uG\pqty{\mu(M),\sigma_0}}{5.3\times10^{-16}}},
}
with
\bme{
    f(\xi)=\frac{1}{2}\xi(\xi^2-3)\pqty{\erf\bqty{\frac{1}{2}\sqrt{\frac{5}{2}}\xi}+\erf\bqty{\sqrt{\frac{5}{2}}\xi}} \\
    +\sqrt{\frac{2}{5\pi}}\Bqty{\pqty{\frac{8}{5}+\frac{31}{4}\xi^2}\exp\bqty{-\frac{5}{8}\xi^2}+\pqty{-\frac{8}{5}+\frac{1}{2}\xi^2}\exp\bqty{-\frac{5}{2}\xi^2}},
}
and the Gaussian distribution $P_\uG(x,\sigma)=\frac{1}{\sqrt{2\pi\sigma^2}}\ee^{-x^2/(2\sigma^2)}$.
We adopt the current observational value of the dark matter density $\Omega_\DM h^2\simeq0.12$~\cite{Planck:2018vyg}. 
In Fig.~\ref{fig: fPBH}, we show the PBH mass spectra with tuned $\sigma_0$ such that the total PBH abundance $f_\PBH^\tot=\int f_\PBH(M)\dd{\ln M}$ becomes unity (left), and also this total abundance $f_\PBH^\tot$ as a function of $\sigma_0^2$ for $\fNL=-1$, $0$, and $1$ (right).

\section{Summary and conclusions}
\label{sec: Summary and conclusions}

In this work, we performed numerical simulations of PBH formation, introducing the local-type non-Gaussianity parametrised by $\fNL$ to the curvature fluctuation for a monochromatic power spectrum on the FLRW universe filled by radiation fluid. 
We have contrasted the results of our numerical simulations with the averaged compaction function approach~\cite{Escriva:2019phb}. In particular, we have found the existence of PBH formation (type I) even for $-1.2 \lesssim \fNL \lesssim -0.336$, in
contrast to the average one which found no type I PBH in this regime~\cite{Kitajima:2021fpq} with the universal threshold $\bar{\calC}_\uc=2/5$~\cite{Escriva:2019phb}.

Our numerical results hence show that for the model we have considered with $\fNL \lesssim -0.336$, the averaged critical compaction function is not equal to $\bar{\mathcal{C}}_{c}=2/5$. It seems to suggest that, though the universality (independent on the profile) of the averaged critical compaction function $\bar{\mathcal{C}}_{c}=2/5$ is basically useful to estimate the PBH formation for a variety of profiles, it could fail for some specific and non-well behaved profiles such as some of the ones we have considered. On the other hand, the analytic estimation of the threshold values through the $q$-parameter formula~\eqref{threshold_anali} has shown to be more robust in this aspect, at least in our model. 
Finally, we have also updated the estimation of the PBH abundance based on the peak theory procedure used in Ref.~\cite{Kitajima:2021fpq}, considering the newly available region for PBH production $-1.2 \lesssim \fNL \lesssim -0.336$ for the model considered.

\acknowledgments

This work is supported by JSPS KAKENHI Grant Numbers
JP19K14707 (Y.T.), JP21K13918 (Y.T.), JP20H01932 (S.Y.), JP20K03968 (S.Y.),
JP19H01895 (C.Y.), JP20H05850 (C.Y.), and JP20H05853 (C.Y.).
A.E. is supported by a postdoctoral grant at the ULB (Université Libre de Bruxelles) University.


\appendix

\bibliographystyle{JHEP}
\bibliography{refs_solved_duplicates.bib}

\providecommand{\href}[2]{#2}\begingroup\raggedright\begin{thebibliography}{10}

\bibitem{Kitajima:2021fpq}
N.~Kitajima, Y.~Tada, S.~Yokoyama and C.-M.~Yoo, \emph{{Primordial black holes
  in peak theory with a non-Gaussian tail}},
  \href{https://doi.org/10.1088/1475-7516/2021/10/053}{\emph{JCAP} {\bfseries
  10} (2021) 053} [\href{https://arxiv.org/abs/2109.00791}{{\ttfamily
  2109.00791}}].

\bibitem{hawking1}
B.J.~Carr and S.W.~Hawking, \emph{{Black holes in the early Universe}},
  {\emph{Mon. Not. Roy. Astron. Soc.} {\bfseries 168} (1974) 399}.

\bibitem{hawking2}
S.~Hawking, \emph{{Gravitationally collapsed objects of very low mass}},
  {\emph{Mon. Not. Roy. Astron. Soc.} {\bfseries 152} (1971) 75}.

\bibitem{acreation1}
Y.B..N.~Zel'dovich, I.~D., \emph{{The Hypothesis of Cores Retarded during
  Expansion and the Hot Cosmological Model}}, {\emph{Soviet Astron. AJ (Engl.
  Transl. ),} {\bfseries 10} (1967) 602}.

\bibitem{LIGO}
{\scshape LIGO Scientific, Virgo} collaboration, \emph{{Observation of
  Gravitational Waves from a Binary Black Hole Merger}},
  \href{https://doi.org/10.1103/PhysRevLett.116.061102}{\emph{Phys. Rev. Lett.}
  {\bfseries 116} (2016) 061102}
  [\href{https://arxiv.org/abs/1602.03837}{{\ttfamily 1602.03837}}].

\bibitem{Bird}
S.~Bird, I.~Cholis, J.B.~Mu\~noz, Y.~Ali-Ha\"\i{}moud, M.~Kamionkowski,
  E.D.~Kovetz et~al., \emph{{Did LIGO detect dark matter?}},
  \href{https://doi.org/10.1103/PhysRevLett.116.201301}{\emph{Phys. Rev. Lett.}
  {\bfseries 116} (2016) 201301}
  [\href{https://arxiv.org/abs/1603.00464}{{\ttfamily 1603.00464}}].

\bibitem{Clesse:2016vqa}
S.~Clesse and J.~Garc\'\i{}a-Bellido, \emph{{The clustering of massive
  Primordial Black Holes as Dark Matter: measuring their mass distribution with
  Advanced LIGO}},
  \href{https://doi.org/10.1016/j.dark.2016.10.002}{\emph{Phys. Dark Univ.}
  {\bfseries 15} (2017) 142}
  [\href{https://arxiv.org/abs/1603.05234}{{\ttfamily 1603.05234}}].

\bibitem{Sasaki:2016jop}
M.~Sasaki, T.~Suyama, T.~Tanaka and S.~Yokoyama, \emph{{Primordial Black Hole
  Scenario for the Gravitational-Wave Event GW150914}},
  \href{https://doi.org/10.1103/PhysRevLett.117.061101}{\emph{Phys. Rev. Lett.}
  {\bfseries 117} (2016) 061101}
  [\href{https://arxiv.org/abs/1603.08338}{{\ttfamily 1603.08338}}].

\bibitem{Carr1}
B.~Carr, F.~Kuhnel and M.~Sandstad, \emph{{Primordial Black Holes as Dark
  Matter}}, \href{https://doi.org/10.1103/PhysRevD.94.083504}{\emph{Phys. Rev.
  D} {\bfseries 94} (2016) 083504}
  [\href{https://arxiv.org/abs/1607.06077}{{\ttfamily 1607.06077}}].

\bibitem{darkmatter1}
J.~Garcia-Bellido, A.D.~Linde and D.~Wands, \emph{{Density perturbations and
  black hole formation in hybrid inflation}},
  \href{https://doi.org/10.1103/PhysRevD.54.6040}{\emph{Phys. Rev. D}
  {\bfseries 54} (1996) 6040}
  [\href{https://arxiv.org/abs/astro-ph/9605094}{{\ttfamily
  astro-ph/9605094}}].

\bibitem{darkmatter2}
M.Y.~Khlopov, \emph{{Primordial Black Holes}},
  \href{https://doi.org/10.1088/1674-4527/10/6/001}{\emph{Res. Astron.
  Astrophys.} {\bfseries 10} (2010) 495}
  [\href{https://arxiv.org/abs/0801.0116}{{\ttfamily 0801.0116}}].

\bibitem{darkmatter3}
M.~Sasaki, T.~Suyama, T.~Tanaka and S.~Yokoyama, \emph{{Primordial black
  holes\textemdash{}perspectives in gravitational wave astronomy}},
  \href{https://doi.org/10.1088/1361-6382/aaa7b4}{\emph{Class. Quant. Grav.}
  {\bfseries 35} (2018) 063001}
  [\href{https://arxiv.org/abs/1801.05235}{{\ttfamily 1801.05235}}].

\bibitem{darkmatter4}
K.~Inomata, M.~Kawasaki, K.~Mukaida, Y.~Tada and T.T.~Yanagida,
  \emph{{Inflationary Primordial Black Holes as All Dark Matter}},
  \href{https://doi.org/10.1103/PhysRevD.96.043504}{\emph{Phys. Rev. D}
  {\bfseries 96} (2017) 043504}
  [\href{https://arxiv.org/abs/1701.02544}{{\ttfamily 1701.02544}}].

\bibitem{darkmatter5}
J.~Georg and S.~Watson, \emph{{A Preferred Mass Range for Primordial Black Hole
  Formation and Black Holes as Dark Matter Revisited}},
  \href{https://doi.org/10.1007/JHEP09(2017)138}{\emph{JHEP} {\bfseries 09}
  (2017) 138} [\href{https://arxiv.org/abs/1703.04825}{{\ttfamily
  1703.04825}}].

\bibitem{darkmatter6}
B.~Carr and J.~Silk, \emph{{Primordial Black Holes as Generators of Cosmic
  Structures}}, \href{https://doi.org/10.1093/mnras/sty1204}{\emph{Mon. Not.
  Roy. Astron. Soc.} {\bfseries 478} (2018) 3756}
  [\href{https://arxiv.org/abs/1801.00672}{{\ttfamily 1801.00672}}].

\bibitem{darkmatter8}
A.~Kashlinsky et~al., \emph{{Electromagnetic probes of primordial black holes
  as dark matter}},  \href{https://arxiv.org/abs/1903.04424}{{\ttfamily
  1903.04424}}.

\bibitem{Clesse:2015wea}
S.~Clesse and J.~Garc\'\i{}a-Bellido, \emph{{Massive Primordial Black Holes
  from Hybrid Inflation as Dark Matter and the seeds of Galaxies}},
  \href{https://doi.org/10.1103/PhysRevD.92.023524}{\emph{Phys. Rev. D}
  {\bfseries 92} (2015) 023524}
  [\href{https://arxiv.org/abs/1501.07565}{{\ttfamily 1501.07565}}].

\bibitem{Clesse:2018ogk}
S.~Clesse, J.~Garc\'\i{}a-Bellido and S.~Orani, \emph{{Detecting the Stochastic
  Gravitational Wave Background from Primordial Black Hole Formation}},
  \href{https://arxiv.org/abs/1812.11011}{{\ttfamily 1812.11011}}.

\bibitem{Tada:2019amh}
Y.~Tada and S.~Yokoyama, \emph{{Primordial black hole tower: Dark matter,
  earth-mass, and LIGO black holes}},
  \href{https://doi.org/10.1103/PhysRevD.100.023537}{\emph{Phys. Rev. D}
  {\bfseries 100} (2019) 023537}
  [\href{https://arxiv.org/abs/1904.10298}{{\ttfamily 1904.10298}}].

\bibitem{Atal:2020yic}
V.~Atal, A.~Sanglas and N.~Triantafyllou, \emph{{NANOGrav signal as mergers of
  Stupendously Large Primordial Black Holes}},
  \href{https://doi.org/10.1088/1475-7516/2021/06/022}{\emph{JCAP} {\bfseries
  06} (2021) 022} [\href{https://arxiv.org/abs/2012.14721}{{\ttfamily
  2012.14721}}].

\bibitem{Atal:2020igj}
V.~Atal, A.~Sanglas and N.~Triantafyllou, \emph{{LIGO/Virgo black holes and
  dark matter: The effect of spatial clustering}},
  \href{https://doi.org/10.1088/1475-7516/2020/11/036}{\emph{JCAP} {\bfseries
  11} (2020) 036} [\href{https://arxiv.org/abs/2007.07212}{{\ttfamily
  2007.07212}}].

\bibitem{Bartolo:2018rku}
N.~Bartolo, V.~De~Luca, G.~Franciolini, M.~Peloso, D.~Racco and A.~Riotto,
  \emph{{Testing primordial black holes as dark matter with LISA}},
  \href{https://doi.org/10.1103/PhysRevD.99.103521}{\emph{Phys. Rev. D}
  {\bfseries 99} (2019) 103521}
  [\href{https://arxiv.org/abs/1810.12224}{{\ttfamily 1810.12224}}].

\bibitem{Clesse:2017bsw}
S.~Clesse and J.~Garc\'\i{}a-Bellido, \emph{{Seven Hints for Primordial Black
  Hole Dark Matter}},
  \href{https://doi.org/10.1016/j.dark.2018.08.004}{\emph{Phys. Dark Univ.}
  {\bfseries 22} (2018) 137}
  [\href{https://arxiv.org/abs/1711.10458}{{\ttfamily 1711.10458}}].

\bibitem{Ezquiaga:2019ftu}
J.M.~Ezquiaga, J.~Garc\'\i{}a-Bellido and V.~Vennin, \emph{{The exponential
  tail of inflationary fluctuations: consequences for primordial black holes}},
  \href{https://doi.org/10.1088/1475-7516/2020/03/029}{\emph{JCAP} {\bfseries
  03} (2020) 029} [\href{https://arxiv.org/abs/1912.05399}{{\ttfamily
  1912.05399}}].

\bibitem{Passaglia:2021jla}
S.~Passaglia and M.~Sasaki, \emph{{Primordial Black Holes from CDM
  Isocurvature}},  \href{https://arxiv.org/abs/2109.12824}{{\ttfamily
  2109.12824}}.

\bibitem{Yoo:2021fxs}
C.-M.~Yoo, T.~Harada, S.~Hirano, H.~Okawa and M.~Sasaki, \emph{{Primordial
  black hole formation from massless scalar isocurvature}},
  \href{https://arxiv.org/abs/2112.12335}{{\ttfamily 2112.12335}}.

\bibitem{Harada:2016mhb}
T.~Harada, C.-M.~Yoo, K.~Kohri, K.-i.~Nakao and S.~Jhingan, \emph{{Primordial
  black hole formation in the matter-dominated phase of the Universe}},
  \href{https://doi.org/10.3847/1538-4357/833/1/61}{\emph{Astrophys. J.}
  {\bfseries 833} (2016) 61}
  [\href{https://arxiv.org/abs/1609.01588}{{\ttfamily 1609.01588}}].

\bibitem{carr75}
B.J.~Carr, \emph{{The Primordial black hole mass spectrum}},
  \href{https://doi.org/10.1086/153853}{\emph{Astrophys. J.} {\bfseries 201}
  (1975) 1}.

\bibitem{musco2005}
I.~Musco, J.C.~Miller and L.~Rezzolla, \emph{{Computations of primordial black
  hole formation}},
  \href{https://doi.org/10.1088/0264-9381/22/7/013}{\emph{Class. Quant. Grav.}
  {\bfseries 22} (2005) 1405}
  [\href{https://arxiv.org/abs/gr-qc/0412063}{{\ttfamily gr-qc/0412063}}].

\bibitem{hawke2002}
I.~Hawke and J.M.~Stewart, \emph{{The dynamics of primordial black hole
  formation}}, \href{https://doi.org/10.1088/0264-9381/19/14/310}{\emph{Class.
  Quant. Grav.} {\bfseries 19} (2002) 3687}.

\bibitem{Harada:2015yda}
T.~Harada, C.-M.~Yoo, T.~Nakama and Y.~Koga, \emph{{Cosmological
  long-wavelength solutions and primordial black hole formation}},
  \href{https://doi.org/10.1103/PhysRevD.91.084057}{\emph{Phys. Rev. D}
  {\bfseries 91} (2015) 084057}
  [\href{https://arxiv.org/abs/1503.03934}{{\ttfamily 1503.03934}}].

\bibitem{Niemeyer2}
J.C.~Niemeyer and K.~Jedamzik, \emph{{Dynamics of primordial black hole
  formation}}, \href{https://doi.org/10.1103/PhysRevD.59.124013}{\emph{Phys.
  Rev. D} {\bfseries 59} (1999) 124013}
  [\href{https://arxiv.org/abs/astro-ph/9901292}{{\ttfamily
  astro-ph/9901292}}].

\bibitem{Shibata:1999zs}
M.~Shibata and M.~Sasaki, \emph{{Black hole formation in the Friedmann
  universe: Formulation and computation in numerical relativity}},
  \href{https://doi.org/10.1103/PhysRevD.60.084002}{\emph{Phys. Rev. D}
  {\bfseries 60} (1999) 084002}
  [\href{https://arxiv.org/abs/gr-qc/9905064}{{\ttfamily gr-qc/9905064}}].

\bibitem{Nakama:2014fra}
T.~Nakama, \emph{{The double formation of primordial black holes}},
  \href{https://doi.org/10.1088/1475-7516/2014/10/040}{\emph{JCAP} {\bfseries
  10} (2014) 040} [\href{https://arxiv.org/abs/1408.0955}{{\ttfamily
  1408.0955}}].

\bibitem{Musco:2018rwt}
I.~Musco, \emph{{Threshold for primordial black holes: Dependence on the shape
  of the cosmological perturbations}},
  \href{https://doi.org/10.1103/PhysRevD.100.123524}{\emph{Phys. Rev. D}
  {\bfseries 100} (2019) 123524}
  [\href{https://arxiv.org/abs/1809.02127}{{\ttfamily 1809.02127}}].

\bibitem{Escriva:2019nsa}
A.~Escriv\`a, \emph{{Simulation of primordial black hole formation using
  pseudo-spectral methods}},
  \href{https://doi.org/10.1016/j.dark.2020.100466}{\emph{Phys. Dark Univ.}
  {\bfseries 27} (2020) 100466}
  [\href{https://arxiv.org/abs/1907.13065}{{\ttfamily 1907.13065}}].

\bibitem{Nakama_2014}
T.~Nakama, T.~Harada, A.G.~Polnarev and J.~Yokoyama, \emph{{Identifying the
  most crucial parameters of the initial curvature profile for primordial black
  hole formation}},
  \href{https://doi.org/10.1088/1475-7516/2014/01/037}{\emph{JCAP} {\bfseries
  01} (2014) 037} [\href{https://arxiv.org/abs/1310.3007}{{\ttfamily
  1310.3007}}].

\bibitem{Escriva:2021aeh}
A.~Escriv\`a, \emph{{PBH formation from spherically symmetric hydrodynamical
  perturbations: a review}},
  \href{https://doi.org/10.3390/universe8020066}{\emph{Universe} {\bfseries 8}
  (2022) 66} [\href{https://arxiv.org/abs/2111.12693}{{\ttfamily 2111.12693}}].

\bibitem{harada}
T.~Harada, C.-M.~Yoo and K.~Kohri, \emph{{Threshold of primordial black hole
  formation}}, \href{https://doi.org/10.1103/PhysRevD.88.084051}{\emph{Phys.
  Rev. D} {\bfseries 88} (2013) 084051}
  [\href{https://arxiv.org/abs/1309.4201}{{\ttfamily 1309.4201}}].

\bibitem{Escriva:2020tak}
A.~Escriv\`a, C.~Germani and R.K.~Sheth, \emph{{Analytical thresholds for black
  hole formation in general cosmological backgrounds}},
  \href{https://doi.org/10.1088/1475-7516/2021/01/030}{\emph{JCAP} {\bfseries
  01} (2021) 030} [\href{https://arxiv.org/abs/2007.05564}{{\ttfamily
  2007.05564}}].

\bibitem{Escriva:2019phb}
A.~Escriv\`a, C.~Germani and R.K.~Sheth, \emph{{Universal threshold for
  primordial black hole formation}},
  \href{https://doi.org/10.1103/PhysRevD.101.044022}{\emph{Phys. Rev. D}
  {\bfseries 101} (2020) 044022}
  [\href{https://arxiv.org/abs/1907.13311}{{\ttfamily 1907.13311}}].

\bibitem{nonlinear}
C.~Germani and R.K.~Sheth, \emph{Nonlinear statistics of primordial black holes
  from gaussian curvature perturbations},
  \href{https://doi.org/10.1103/PhysRevD.101.063520}{\emph{Phys. Rev. D}
  {\bfseries 101} (2020) 063520}.

\bibitem{DeLuca:2019qsy}
V.~De~Luca, G.~Franciolini, A.~Kehagias, M.~Peloso, A.~Riotto and C.~\"Unal,
  \emph{{The Ineludible non-Gaussianity of the Primordial Black Hole
  Abundance}}, \href{https://doi.org/10.1088/1475-7516/2019/07/048}{\emph{JCAP}
  {\bfseries 07} (2019) 048}
  [\href{https://arxiv.org/abs/1904.00970}{{\ttfamily 1904.00970}}].

\bibitem{newnicolla}
A.~Kalaja, N.~Bellomo, N.~Bartolo, D.~Bertacca, S.~Matarrese, I.~Musco et~al.,
  \emph{{From Primordial Black Holes Abundance to Primordial Curvature Power
  Spectrum (and back)}},
  \href{https://doi.org/10.1088/1475-7516/2019/10/031}{\emph{JCAP} {\bfseries
  10} (2019) 031} [\href{https://arxiv.org/abs/1908.03596}{{\ttfamily
  1908.03596}}].

\bibitem{Erfani:2021rmw}
E.~Erfani, H.~Kameli and S.~Baghram, \emph{{Primordial black holes in the
  excursion set theory}},
  \href{https://doi.org/10.1093/mnras/stab1403}{\emph{Mon. Not. Roy. Astron.
  Soc.} {\bfseries 505} (2021) 1787}
  [\href{https://arxiv.org/abs/2101.07812}{{\ttfamily 2101.07812}}].

\bibitem{Wu:2020ilx}
Y.-P.~Wu, \emph{{Peak statistics for the primordial black hole abundance}},
  \href{https://doi.org/10.1016/j.dark.2020.100654}{\emph{Phys. Dark Univ.}
  {\bfseries 30} (2020) 100654}
  [\href{https://arxiv.org/abs/2005.00441}{{\ttfamily 2005.00441}}].

\bibitem{DeLuca:2020ioi}
V.~De~Luca, G.~Franciolini and A.~Riotto, \emph{{On the Primordial Black Hole
  Mass Function for Broad Spectra}},
  \href{https://doi.org/10.1016/j.physletb.2020.135550}{\emph{Phys. Lett. B}
  {\bfseries 807} (2020) 135550}
  [\href{https://arxiv.org/abs/2001.04371}{{\ttfamily 2001.04371}}].

\bibitem{Young:2020xmk}
S.~Young and M.~Musso, \emph{{Application of peaks theory to the abundance of
  primordial black holes}},
  \href{https://doi.org/10.1088/1475-7516/2020/11/022}{\emph{JCAP} {\bfseries
  11} (2020) 022} [\href{https://arxiv.org/abs/2001.06469}{{\ttfamily
  2001.06469}}].

\bibitem{Yoo:2020dkz}
C.-M.~Yoo, T.~Harada, S.~Hirano and K.~Kohri, \emph{{Abundance of Primordial
  Black Holes in Peak Theory for an Arbitrary Power Spectrum}},
  \href{https://doi.org/10.1093/ptep/ptaa155}{\emph{PTEP} {\bfseries 2021}
  (2021) 013E02} [\href{https://arxiv.org/abs/2008.02425}{{\ttfamily
  2008.02425}}].

\bibitem{yoo}
C.-M.~Yoo, J.-O.~Gong and S.~Yokoyama, \emph{{Abundance of primordial black
  holes with local non-Gaussianity in peak theory}},
  \href{https://doi.org/10.1088/1475-7516/2019/09/033}{\emph{JCAP} {\bfseries
  09} (2019) 033} [\href{https://arxiv.org/abs/1906.06790}{{\ttfamily
  1906.06790}}].

\bibitem{Gow:2020bzo}
A.D.~Gow, C.T.~Byrnes, P.S.~Cole and S.~Young, \emph{{The power spectrum on
  small scales: Robust constraints and comparing PBH methodologies}},
  \href{https://doi.org/10.1088/1475-7516/2021/02/002}{\emph{JCAP} {\bfseries
  02} (2021) 002} [\href{https://arxiv.org/abs/2008.03289}{{\ttfamily
  2008.03289}}].

\bibitem{Young:2019osy}
S.~Young, \emph{{The primordial black hole formation criterion re-examined:
  Parametrisation, timing and the choice of window function}},
  \href{https://doi.org/10.1142/S0218271820300025}{\emph{Int. J. Mod. Phys. D}
  {\bfseries 29} (2019) 2030002}
  [\href{https://arxiv.org/abs/1905.01230}{{\ttfamily 1905.01230}}].

\bibitem{Young:2019yug}
S.~Young, I.~Musco and C.T.~Byrnes, \emph{{Primordial black hole formation and
  abundance: contribution from the non-linear relation between the density and
  curvature perturbation}},
  \href{https://doi.org/10.1088/1475-7516/2019/11/012}{\emph{JCAP} {\bfseries
  11} (2019) 012} [\href{https://arxiv.org/abs/1904.00984}{{\ttfamily
  1904.00984}}].

\bibitem{Young:2013oia}
S.~Young and C.T.~Byrnes, \emph{{Primordial black holes in non-Gaussian
  regimes}}, \href{https://doi.org/10.1088/1475-7516/2013/08/052}{\emph{JCAP}
  {\bfseries 08} (2013) 052} [\href{https://arxiv.org/abs/1307.4995}{{\ttfamily
  1307.4995}}].

\bibitem{Young:2014ana}
S.~Young, C.T.~Byrnes and M.~Sasaki, \emph{{Calculating the mass fraction of
  primordial black holes}},
  \href{https://doi.org/10.1088/1475-7516/2014/07/045}{\emph{JCAP} {\bfseries
  07} (2014) 045} [\href{https://arxiv.org/abs/1405.7023}{{\ttfamily
  1405.7023}}].

\bibitem{Germani:2018jgr}
C.~Germani and I.~Musco, \emph{{Abundance of Primordial Black Holes Depends on
  the Shape of the Inflationary Power Spectrum}},
  \href{https://doi.org/10.1103/PhysRevLett.122.141302}{\emph{Phys. Rev. Lett.}
  {\bfseries 122} (2019) 141302}
  [\href{https://arxiv.org/abs/1805.04087}{{\ttfamily 1805.04087}}].

\bibitem{Garriga}
C.-M.~Yoo, T.~Harada, J.~Garriga and K.~Kohri, \emph{{Primordial black hole
  abundance from random Gaussian curvature perturbations and a local density
  threshold}}, \href{https://doi.org/10.1093/ptep/pty120}{\emph{PTEP}
  {\bfseries 2018} (2018) 123E01}
  [\href{https://arxiv.org/abs/1805.03946}{{\ttfamily 1805.03946}}].

\bibitem{Suyama:2019npc}
T.~Suyama and S.~Yokoyama, \emph{{A novel formulation of the primordial black
  hole mass function}}, \href{https://doi.org/10.1093/ptep/ptaa011}{\emph{PTEP}
  {\bfseries 2020} (2020) 023E03}
  [\href{https://arxiv.org/abs/1912.04687}{{\ttfamily 1912.04687}}].

\bibitem{Ando:2018qdb}
K.~Ando, K.~Inomata and M.~Kawasaki, \emph{{Primordial black holes and
  uncertainties in the choice of the window function}},
  \href{https://doi.org/10.1103/PhysRevD.97.103528}{\emph{Phys. Rev. D}
  {\bfseries 97} (2018) 103528}
  [\href{https://arxiv.org/abs/1802.06393}{{\ttfamily 1802.06393}}].

\bibitem{Zaballa:2006kh}
I.~Zaballa, A.M.~Green, K.A.~Malik and M.~Sasaki, \emph{{Constraints on the
  primordial curvature perturbation from primordial black holes}},
  \href{https://doi.org/10.1088/1475-7516/2007/03/010}{\emph{JCAP} {\bfseries
  03} (2007) 010} [\href{https://arxiv.org/abs/astro-ph/0612379}{{\ttfamily
  astro-ph/0612379}}].

\bibitem{Yokoyama:1998xd}
J.~Yokoyama, \emph{{Cosmological constraints on primordial black holes produced
  in the near critical gravitational collapse}},
  \href{https://doi.org/10.1103/PhysRevD.58.107502}{\emph{Phys. Rev. D}
  {\bfseries 58} (1998) 107502}
  [\href{https://arxiv.org/abs/gr-qc/9804041}{{\ttfamily gr-qc/9804041}}].

\bibitem{Tada:2021zzj}
Y.~Tada and V.~Vennin, \emph{{Statistics of coarse-grained cosmological fields
  in stochastic inflation}},
  \href{https://arxiv.org/abs/2111.15280}{{\ttfamily 2111.15280}}.

\bibitem{Atal:2018neu}
V.~Atal and C.~Germani, \emph{{The role of non-gaussianities in Primordial
  Black Hole formation}},
  \href{https://doi.org/10.1016/j.dark.2019.100275}{\emph{Phys. Dark Univ.}
  {\bfseries 24} (2019) 100275}
  [\href{https://arxiv.org/abs/1811.07857}{{\ttfamily 1811.07857}}].

\bibitem{vicente-garriga}
V.~Atal, J.~Garriga and A.~Marcos-Caballero, \emph{{Primordial black hole
  formation with non-Gaussian curvature perturbations}},
  \href{https://doi.org/10.1088/1475-7516/2019/09/073}{\emph{JCAP} {\bfseries
  09} (2019) 073} [\href{https://arxiv.org/abs/1905.13202}{{\ttfamily
  1905.13202}}].

\bibitem{Hayato2}
S.~Passaglia, W.~Hu and H.~Motohashi, \emph{{Primordial black holes and local
  non-Gaussianity in canonical inflation}},
  \href{https://doi.org/10.1103/PhysRevD.99.043536}{\emph{Phys. Rev. D}
  {\bfseries 99} (2019) 043536}
  [\href{https://arxiv.org/abs/1812.08243}{{\ttfamily 1812.08243}}].

\bibitem{Cai:2017bxr}
Y.-F.~Cai, X.~Chen, M.H.~Namjoo, M.~Sasaki, D.-G.~Wang and Z.~Wang,
  \emph{{Revisiting non-Gaussianity from non-attractor inflation models}},
  \href{https://doi.org/10.1088/1475-7516/2018/05/012}{\emph{JCAP} {\bfseries
  05} (2018) 012} [\href{https://arxiv.org/abs/1712.09998}{{\ttfamily
  1712.09998}}].

\bibitem{Bullock:1996at}
J.S.~Bullock and J.R.~Primack, \emph{{NonGaussian fluctuations and primordial
  black holes from inflation}},
  \href{https://doi.org/10.1103/PhysRevD.55.7423}{\emph{Phys. Rev. D}
  {\bfseries 55} (1997) 7423}
  [\href{https://arxiv.org/abs/astro-ph/9611106}{{\ttfamily
  astro-ph/9611106}}].

\bibitem{Pattison_2017}
C.~Pattison, V.~Vennin, H.~Assadullahi and D.~Wands, \emph{{Quantum diffusion
  during inflation and primordial black holes}},
  \href{https://doi.org/10.1088/1475-7516/2017/10/046}{\emph{JCAP} {\bfseries
  10} (2017) 046} [\href{https://arxiv.org/abs/1707.00537}{{\ttfamily
  1707.00537}}].

\bibitem{PinaAvelino:2005rm}
P.~Pina~Avelino, \emph{{Primordial black hole constraints on non-gaussian
  inflation models}},
  \href{https://doi.org/10.1103/PhysRevD.72.124004}{\emph{Phys. Rev. D}
  {\bfseries 72} (2005) 124004}
  [\href{https://arxiv.org/abs/astro-ph/0510052}{{\ttfamily
  astro-ph/0510052}}].

\bibitem{Young:2014oea}
S.~Young and C.T.~Byrnes, \emph{{Long-short wavelength mode coupling tightens
  primordial black hole constraints}},
  \href{https://doi.org/10.1103/PhysRevD.91.083521}{\emph{Phys. Rev. D}
  {\bfseries 91} (2015) 083521}
  [\href{https://arxiv.org/abs/1411.4620}{{\ttfamily 1411.4620}}].

\bibitem{Riccardi:2021rlf}
F.~Riccardi, M.~Taoso and A.~Urbano, \emph{{Solving peak theory in the presence
  of local non-gaussianities}},
  \href{https://doi.org/10.1088/1475-7516/2021/08/060}{\emph{JCAP} {\bfseries
  08} (2021) 060} [\href{https://arxiv.org/abs/2102.04084}{{\ttfamily
  2102.04084}}].

\bibitem{Young:2015cyn}
S.~Young, D.~Regan and C.T.~Byrnes, \emph{{Influence of large local and
  non-local bispectra on primordial black hole abundance}},
  \href{https://doi.org/10.1088/1475-7516/2016/02/029}{\emph{JCAP} {\bfseries
  02} (2016) 029} [\href{https://arxiv.org/abs/1512.07224}{{\ttfamily
  1512.07224}}].

\bibitem{Hidalgo:2007vk}
J.C.~Hidalgo, \emph{{The effect of non-Gaussian curvature perturbations on the
  formation of primordial black holes}},
  \href{https://arxiv.org/abs/0708.3875}{{\ttfamily 0708.3875}}.

\bibitem{Atal:2021jyo}
V.~Atal and G.~Dom\`enech, \emph{{Probing non-Gaussianities with the high
  frequency tail of induced gravitational waves}},
  \href{https://doi.org/10.1088/1475-7516/2021/06/001}{\emph{JCAP} {\bfseries
  06} (2021) 001} [\href{https://arxiv.org/abs/2103.01056}{{\ttfamily
  2103.01056}}].

\bibitem{Davies:2021loj}
M.W.~Davies, P.~Carrilho and D.J.~Mulryne, \emph{{Non-Gaussianity in
  inflationary scenarios for primordial black holes}},
  \href{https://arxiv.org/abs/2110.08189}{{\ttfamily 2110.08189}}.

\bibitem{Taoso:2021uvl}
M.~Taoso and A.~Urbano, \emph{{Non-gaussianities for primordial black hole
  formation}}, \href{https://doi.org/10.1088/1475-7516/2021/08/016}{\emph{JCAP}
  {\bfseries 08} (2021) 016}
  [\href{https://arxiv.org/abs/2102.03610}{{\ttfamily 2102.03610}}].

\bibitem{Young:2022phe}
S.~Young, \emph{{Peaks and primordial black holes: the effect of
  non-Gaussianity}},  \href{https://arxiv.org/abs/2201.13345}{{\ttfamily
  2201.13345}}.

\bibitem{Kopp:2010sh}
M.~Kopp, S.~Hofmann and J.~Weller, \emph{{Separate Universes Do Not Constrain
  Primordial Black Hole Formation}},
  \href{https://doi.org/10.1103/PhysRevD.83.124025}{\emph{Phys. Rev. D}
  {\bfseries 83} (2011) 124025}
  [\href{https://arxiv.org/abs/1012.4369}{{\ttfamily 1012.4369}}].

\bibitem{Atal:2019erb}
V.~Atal, J.~Cid, A.~Escriv\`a and J.~Garriga, \emph{{PBH in single field
  inflation: the effect of shape dispersion and non-Gaussianities}},
  \href{https://doi.org/10.1088/1475-7516/2020/05/022}{\emph{JCAP} {\bfseries
  05} (2020) 022} [\href{https://arxiv.org/abs/1908.11357}{{\ttfamily
  1908.11357}}].

\bibitem{Bardeen:1985tr}
J.M.~Bardeen, J.R.~Bond, N.~Kaiser and A.S.~Szalay, \emph{{The Statistics of
  Peaks of Gaussian Random Fields}},
  \href{https://doi.org/10.1086/164143}{\emph{Astrophys. J.} {\bfseries 304}
  (1986) 15}.

\bibitem{misnersharp}
C.W.~Misner and D.H.~Sharp, \emph{{Relativistic equations for adiabatic,
  spherically symmetric gravitational collapse}},
  \href{https://doi.org/10.1103/PhysRev.136.B571}{\emph{Phys. Rev.} {\bfseries
  136} (1964) B571}.

\bibitem{Polnarev:2006aa}
A.G.~Polnarev and I.~Musco, \emph{{Curvature profiles as initial conditions for
  primordial black hole formation}},
  \href{https://doi.org/10.1088/0264-9381/24/6/003}{\emph{Class. Quant. Grav.}
  {\bfseries 24} (2007) 1405}
  [\href{https://arxiv.org/abs/gr-qc/0605122}{{\ttfamily gr-qc/0605122}}].

\bibitem{Polnarev:2012bi}
A.G.~Polnarev, T.~Nakama and J.~Yokoyama, \emph{{Self-consistent initial
  conditions for primordial black hole formation}},
  \href{https://doi.org/10.1088/1475-7516/2012/09/027}{\emph{JCAP} {\bfseries
  09} (2012) 027} [\href{https://arxiv.org/abs/1204.6601}{{\ttfamily
  1204.6601}}].

\bibitem{choptuik}
M.W.~Choptuik, \emph{{Universality and scaling in gravitational collapse of a
  massless scalar field}},
  \href{https://doi.org/10.1103/PhysRevLett.70.9}{\emph{Phys. Rev. Lett.}
  {\bfseries 70} (1993) 9}.

\bibitem{coleman}
C.R.~Evans and J.S.~Coleman, \emph{Critical phenomena and self-similarity in
  the gravitational collapse of radiation fluid},
  \href{https://doi.org/10.1103/PhysRevLett.72.1782}{\emph{Phys. Rev. Lett.}
  {\bfseries 72} (1994) 1782}
  [\href{https://arxiv.org/abs/gr-qc/9402041}{{\ttfamily gr-qc/9402041}}].

\bibitem{renormalizationcriticalcollapse}
T.~Koike, T.~Hara and S.~Adachi, \emph{{Critical behavior in gravitational
  collapse of radiation fluid: A Renormalization group (linear perturbation)
  analysis}}, \href{https://doi.org/10.1103/PhysRevLett.74.5170}{\emph{Phys.
  Rev. Lett.} {\bfseries 74} (1995) 5170}
  [\href{https://arxiv.org/abs/gr-qc/9503007}{{\ttfamily gr-qc/9503007}}].

\bibitem{Niemeyer1}
J.C.~Niemeyer and K.~Jedamzik, \emph{{Near-critical gravitational collapse and
  the initial mass function of primordial black holes}},
  \href{https://doi.org/10.1103/PhysRevLett.80.5481}{\emph{Phys. Rev. Lett.}
  {\bfseries 80} (1998) 5481}
  [\href{https://arxiv.org/abs/astro-ph/9709072}{{\ttfamily
  astro-ph/9709072}}].

\bibitem{musco2009}
I.~Musco, J.C.~Miller and A.G.~Polnarev, \emph{{Primordial black hole formation
  in the radiative era: Investigation of the critical nature of the collapse}},
  \href{https://doi.org/10.1088/0264-9381/26/23/235001}{\emph{Class. Quant.
  Grav.} {\bfseries 26} (2009) 235001}
  [\href{https://arxiv.org/abs/0811.1452}{{\ttfamily 0811.1452}}].

\bibitem{Escriva:2021pmf}
A.~Escriv\`a and A.E.~Romano, \emph{{Effects of the shape of curvature peaks on
  the size of primordial black holes}},
  \href{https://doi.org/10.1088/1475-7516/2021/05/066}{\emph{JCAP} {\bfseries
  05} (2021) 066} [\href{https://arxiv.org/abs/2103.03867}{{\ttfamily
  2103.03867}}].

\bibitem{Planck:2018vyg}
{\scshape Planck} collaboration, \emph{{Planck 2018 results. VI. Cosmological
  parameters}},
  \href{https://doi.org/10.1051/0004-6361/201833910}{\emph{Astron. Astrophys.}
  {\bfseries 641} (2020) A6}
  [\href{https://arxiv.org/abs/1807.06209}{{\ttfamily 1807.06209}}].

\end{thebibliography}\endgroup

\end{document}